\documentclass[review]{elsarticle}

\usepackage[colorinlistoftodos]{todonotes}
\usepackage{multirow}
\usepackage{rotating}
\usepackage{lineno,hyperref}
\usepackage{textgreek}
\modulolinenumbers[5]

\journal{Journal of Information Processing and Management}

\begin{document}

\begin{frontmatter}

% \title{Elsevier \LaTeX\ template\tnoteref{mytitlenote}}
\title{Beyond MeSH: Fine-Grained Semantic Indexing of Biomedical Literature 
based on Weak Supervision}
% \tnotetext[mytitlenote]{Fully documented templates are available in the elsarticle package on \href{http://www.ctan.org/tex-archive/macros/latex/contrib/elsarticle}{CTAN}.}

% \address{Radarweg 29, Amsterdam}
\fntext[myfootnote]{\textcopyright 2020. This manuscript version is made available under the CC-BY-NC-ND 4.0 license http://creativecommons.org/licenses/by-nc-nd/4.0/}

% \author{Anastasios Nentidis\fnref{myfootnote}\textsuperscript{1,2}, Anastasia Krithara\textsuperscript{1}, Grigorios Tsoumakas\textsuperscript{2}, Georgios Paliouras\textsuperscript{1}}
\author[NCSRD,AUTH]{Anastasios Nentidis\corref{mycorrespondingauthor}}
\cortext[mycorrespondingauthor]{Corresponding author}
\ead{tasosnent@iit.demokritos.gr}
\author[NCSRD]{Anastasia Krithara}
\author[AUTH]{Grigorios Tsoumakas}
\author[NCSRD]{Georgios Paliouras}
% \author[mymainaddress,mysecondaryaddress]{Elsevier Inc}

% \author[mysecondaryaddress]{Global Customer Service\corref{mycorrespondingauthor}}

\address[NCSRD]{Institute of Informatics and Telecommunications, NCSR Demokritos, Athens, Greece}
\address[AUTH]{School of Informatics, Aristotle University of Thessaloniki, Thessaloniki, Greece}

\begin{abstract}
In this work, we propose a method for the automated refinement of subject annotations in biomedical literature at the level of concepts. 
Semantic indexing and search of biomedical articles in MEDLINE/PubMed are based on semantic subject annotations with MeSH descriptors that may correspond to several related but distinct biomedical concepts.
% Biomedical literature in MEDLINE/PubMed is semantically indexed with MeSH thesaurus entries (subject annotations) that may correspond to more than one related but distinct domain concepts. 
Such semantic annotations do not adhere to the level of detail available in the domain knowledge and may not be sufficient to fulfil the information needs of experts in the domain.
% In such cases, the subject annotations do not follow the level of detail available in the domain and do not always suffice to meet the information needs of domain experts. 
% detail based on the abstract and the title of the article
To this end, we propose a new method that uses weak supervision to train a concept annotator on the literature available for a particular disease. We test this method on the MeSH descriptors for two diseases: Alzheimer's Disease and Duchenne Muscular Dystrophy.
The results indicate that concept-occurrence is a strong heuristic for automated subject annotation refinement and its use as weak supervision can lead to improved concept-level annotations.
The fine-grained semantic annotations can enable more precise literature retrieval, sustain the semantic integration of subject annotations with other domain resources and ease the maintenance of consistent subject annotations, as new more detailed entries are added in the MeSH thesaurus over time.
% during its evolution along with the biomedical domain. 
\end{abstract}

\begin{keyword}
% \texttt{elsarticle.cls}\sep \LaTeX\sep Elsevier \sep template
semantic indexing \sep MeSH \sep biomedical literature \sep weak supervision
\MSC[2010] 00-01\sep  99-00 
\end{keyword}

\end{frontmatter}

% \linenumbers

\section{Introduction}

% \paragraph{Installation} If the document class \emph{elsarticle} is not available on your computer, you can download and install the system package \emph{texlive-publishers} (Linux) or install the \LaTeX\ package \emph{elsarticle} using the package manager of your \TeX\ installation, which is typically \TeX\ Live or Mik\TeX.

% \paragraph{Usage} Once the package is properly installed, you can use the document class \emph{elsarticle} to create a manuscript. Please make sure that your manuscript follows the guidelines in the Guide for Authors of the relevant journal. It is not necessary to typeset your manuscript in exactly the same way as an article, unless you are submitting to a camera-ready copy (CRC) journal.

% \paragraph{Functionality} The Elsevier article class is based on the standard article class and supports almost all of the functionality of that class. In addition, it features commands and options to format the
% \begin{itemize}
% \item document style
% \item baselineskip
% \item front matter
% \item keywords and MSC codes
% \item theorems, definitions and proofs
% \item lables of enumerations
% \item citation style and labeling.
% \end{itemize}

% Retrieval of relevant academic publications is important both directly to biomedical researchers looking for specific information, as well as to many downstream tasks, such as question answering and technologically assisted reviews.
Retrieval of relevant biomedical scientific publications is essential directly to researchers in search of specific information, as well as to a range of downstream tasks, including technologically assisted reviews and question answering.
% As traditional keyword-based systems face important challenges, such as synonymy and polysemy, semantic approaches have been developed that exploit ontological resources to retrieve material relevant to specific domain entities. The annotation of documents as relevant to specific conceptual entities is called semantic indexing and the use of such annotations sor information retrieal is called semantic search.
Semantic approaches in information retrieval confront important challenges of traditional keyword-based search, such as synonymy and polysemy, by exploiting ontological domain resources. The annotation of documents as relevant to conceptual domain entities is called semantic indexing and can be used to support semantic search.
% , that is search based on semantic information

\begin{figure}[!t]
\centerline{\includegraphics[width=0.92\textwidth]{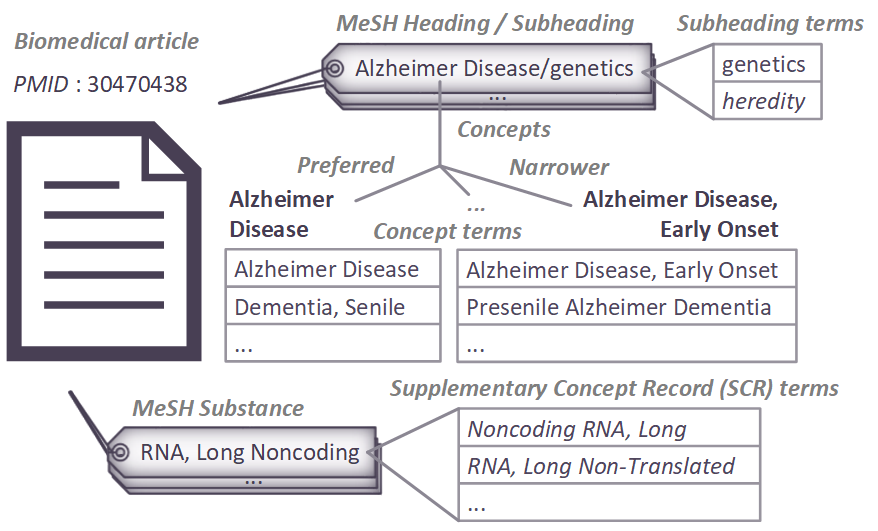}}
\caption{ An article in PubMed/MEDLINE with some of its MeSH annotations. The internal structure of the corresponding MeSH entries is also shown. }\label{fig:MeSH_Example}
\end{figure}

To enable semantic search within PubMed/MEDLINE\footnote{\url{https://www.ncbi.nlm.nih.gov/pubmed/}}, the National Library of Medicine (NLM) of the United States has developed and maintains the Medical Subject Headings (MeSH) thesaurus\footnote{\url{https://www.nlm.nih.gov/mesh/}}. 
MeSH consists of a collection of subject terms grouped into interconnected entries, such as hierarchically organized subject headings or descriptors, topical subheadings or qualifiers, and supplementary concept records.
% MeSH is a set of subject terms organized into inter-related entries, including hierarchically organized subject descriptors, topical qualifiers, also known as subheadings, and supplementary concept records. 
The PubMed/MEDLINE publications are manually annotated with MeSH entries by expert indexers in NLM, as in the example shown in Fig.~\ref{fig:MeSH_Example}.
% NLM indexers manually annotate PubMed/MEDLINE publications with MeSH terms. 
Manually processing the ever-growing volume of publications is a challenge, and for this reason the Medical Text Indexer (MTI) \cite{morkBioasq2014} has been developed in NLM. MTI is a specialised tool to help the indexers by automatically suggesting annotations. 
% As an example, Fig.~\ref{fig:MeSH_Example} shows some of the annotations of an article using MeSH.

Each MeSH descriptor \textit{t} is composed of a group of terms that are considered equivalent for semantic indexing and search.
% , without necessarily being synonymous. 
In particular, each descriptor \textit{t} corresponds to a set of distinct MeSH concepts \textit{C\textsubscript{t}} and each concept \textit{c\textsubscript{i}} in \textit{C\textsubscript{t}} corresponds to a set of synonymous terms. 
One of the concepts \textit{c\textsubscript{i}} in \textit{C\textsubscript{t}} is the preferred concept \textit{c\textsubscript{pref}}, while the rest can be narrower, broader or just related to \textit{c\textsubscript{pref}}. 
Only descriptors \textit{t} where exactly one concept (\textit{c\textsubscript{top}}) in \textit{C\textsubscript{t}} is broader than any other concept in \textit{C\textsubscript{t}} are considered in this work.
% In this work we only consider descriptors where exactly one concept \textit{c\textsubscript{top}} in \textit{C\textsubscript{t}} is broader than any other concept in \textit{C\textsubscript{t}}.
In particular, if the \textit{c\textsubscript{top}} is also the \textit{c\textsubscript{pref}} only narrower concepts should be included in the \textit{C\textsubscript{t}}. Otherwise, the \textit{c\textsubscript{top}} should be broader to the \textit{c\textsubscript{pref}} and the rest concepts in \textit{C\textsubscript{t}} can be narrower or related to \textit{c\textsubscript{pref}} but still narrower to \textit{c\textsubscript{top}}.

For example, the MeSH descriptor for ``Alzheimer Disease'' (AD), shown in Fig.~\ref{fig:MeSH_Example}, is composed by 22 terms organised into seven concepts, namely the \textit{c\textsubscript{pref}} ``Alzheimer Disease'' and six narrower concepts, including ``Alzheimer Disease, Late Onset'' and ``Alzheimer Disease, Early Onset''.
% In the example of Fig.~\ref{fig:MeSH_Example}, the MeSH descriptor for ``Alzheimer Disease'' (AD}) consists of 22 terms from seven distinct concepts, namely the \textit{c\textsubscript{pref}} ``Alzheimer Disease'' and six narrower concepts including ``Alzheimer Disease, Early Onset'' and ``Alzheimer Disease, Late Onset''. 
In this case, \textit{c\textsubscript{pref}} is also the \textit{c\textsubscript{top}}.
It is worth noting that the MeSH concepts discussed here are distinct from the supplementary concept records, which predominantly relate to chemicals still not included in any MeSH descriptor. The supplementary concept records are used separately to annotate articles in the same way as ``RNA, Long Noncoding'' is used in Fig.~\ref{fig:MeSH_Example}.

% In particular, for each descriptor \textit{t} there is a set \textit{C\textsubscript{t}} of all the MeSH concepts \textit{c\textsubscript{i}} corresponding to \textit{t}. 
% For instance, the MeSH descriptor for ``Alzheimer Disease'' (AD) has 22 terms from seven distinct concepts including ``early onset AD'' and ``late onset AD'', currently handled as equivalent for semantic indexing and search. 
% Since these relations are hierarchical, there will always be at least one concept that is broader than any other concept in \textit{C\textsubscript{t}}.
% In the example of Fig.~\ref{fig:MeSH_Example} the homonymous concept ``Alzheimer Disease'' is the preferred concept and the concept of ``early onset AD'' is one of the narrower concepts of the descriptor.
% In this work we only consider descriptors
% where the preferred concept \textit{c\textsubscript{pref}} is the broader of all concepts \textit{c\textsubscript{i}} in \textit{C\textsubscript{t}}.
% where exactly one concept \textit{c\textsubscript{top}} in their \textit{C\textsubscript{t}} is broader than any other concept in \textit{C\textsubscript{t}}. In the case of AD, \textit{c\textsubscript{pref}} is also the \textit{c\textsubscript{top}}.

 Although MeSH contains almost 29,000 headings, it often aggregates into the same descriptor some closely related yet distinct concepts, failing to tackle the fine-grained conceptual facets of biomedical knowledge.
% MeSH does not provide conceptually fine granularity of annotations and groups some closely related but distinct concepts into the same descriptor. 
% Some cases where this kind of grouping is applied are concepts with only few relevant citations, concepts with overlap in their meaning and concepts where the distinction is considered not clear regarding their use for annotation and retrieval of biomedical literature. 
By way of illustration, Fig.~\ref{fig:MeSH_Example} presents how terms of concepts for different types of AD, are aggregated in the same descriptor, under the implicit assumption that their distinction is beyond the anticipated usage of MeSH.
% For example, as shown in Fig.~\ref{fig:MeSH_Example} concept terms for different types of AD are grouped into a single descriptor assuming that their distinction is beyond the intended use of MeSH.
% , as some of the concepts occur in very few articles.
% For example, terms for different species of whales are grouped in one descriptor considering that their distinction is beyond biomedical focus and some of them would correspond to only few articles. Another case of grouping is for concepts with overlap in their meaning such as ``Isometric'' and ``Aerobic'' exercise which are both grouped in the descriptor for ``Exercise''. In addition, even clearly distinguishable concepts are grouped if their application in literature indexing is considered confusing. For example, the process of ``DNA Fingerprinting'' and its product, the ``DNA Fingerprints''\footnote{Examples drawn from https://www.nlm.nih.gov/mesh/meshrels.html}.
% For details and examples consult the website of MeSH\footnote{https://www.nlm.nih.gov/mesh/meshrels.html}.
Therefore, biomedical experts with specialization on particular biomedical domains usually need to investigate at levels of semantic granularity not provided by MeSH descriptors.
% Therefore, experts specializing in a specific biomedical domain typically need to drill down to a level of granularity that is not supported by MeSH descriptors.
This is particularly important for MeSH descriptors of diseases, as segmentation of the relevant literature into fine-grained parts can uncover differences in certain patient types and make detailed information available to precision medicine applications. 

The majority of descriptors in MeSH corresponds to a single concept, yet some descriptors are associated with up to fifty five concepts.
% Most of the descriptors are associated with a single concept, but there are descriptors with up to 55 concepts. 
In MeSH 2018 more than 10,000 descriptors (35\%) have two or more concepts, in which cases  concept-level semantic annotations could be useful. 
% In total there are more than 10,000 descriptors in MeSH 2018 (35\%) with two or more concepts, where fine-grained semantic indexing could be useful. 
Furthermore, during the evolution of the MeSH thesaurus, concepts aggregated into existing descriptors can be detached into new descriptors. 
For instance, at some point a new descriptor can be added for ``Alzheimer Disease, Early Onset'' as a narrower descriptor of the one for AD. 
% In addition, as the MeSH thesaurus evolves, new descriptors are added as children of existing descriptors. For example, a new descriptor can be added for ``Early Onset Alzheimer Disease'' in the future. 
Even though this evolution can eventually lead to more fine-grained indexing with MeSH descriptors, existing semantic annotations based on older versions of the thesaurus will still need revision, so that semantic search can be homogeneous throughout all relevant literature.
% Although this evolution pushes towards more fine-grained indexing, annotations of articles based on previous versions of MeSH need to be revised to provide homogeneous semantic search. 
% Such new descriptors will be particularly difficult for existing semantic annotation systems until manual annotations for a sufficient number of articles become finally available.
% Prediction of such new descriptors by semantic annotation systems will be difficult until a sufficient amount of manual annotations become eventually available.
% Nevertheless, 
Automated fine-grained indexing could facilitate retrospective revision of semantic annotations for such new descriptors. 
% Automated fine-grained indexing could support such revision of semantic annotations for new descriptors.
% Our approach for fine-grained indexing can support such revision of semantic annotations, based on concept-occurrence in the abstract of the articles.

The goal of the work presented here is to achieve fine-grained semantic indexing of biomedical literature, beyond the descriptors of MeSH, at the semantic level of the corresponding concepts.
% Our work aims to achieve automated fine-grained indexing of biomedical literature, beyond MeSH descriptors, focusing on the corresponding MeSH concepts.
Under the lack of ground truth data for semantic indexing at this level of granularity, we examine the occurrence of concept terms in article abstracts as a heuristic.
% As there is no ground truth for such fine-grained indexing, we investigate the occurrence of concept terms in article abstracts as a heuristic.
To this end, we exploit the existing manual annotations available, focusing on articles that have already been annotated with the MeSH descriptor corresponding to a disease. 
% Toward this direction, we take advantage of the existing MeSH annotations, focusing on citations already annotated with a MeSH descriptor for a disease. 
% \todo{To mention the motivation for focusing on diseases, rather than focusing on something else.}
Furthermore, we also focus on narrower concepts that constitute about 85\% of relations among concepts in MeSH 2018, taking advantage of the conceptual structure of the disease descriptors.
% In addition, we also exploit the conceptual structure of each disease descriptor focusing on narrower concepts that account for about 85\% of concept relations in MeSH 2018. 
% \todo{We also have "broader" and "related" in the case of DMD, but it is not our focus, as we "invert them" into two "narrower" relations again.}
Though the proposed method is applicable to any type of descriptor, focusing on descriptors for diseases is a priority, as the narrower concepts correspond to disease types. Identifying literature directly relevant to specific types of a disease can accelerate the understanding of the disease mechanisms, the design of targeted treatments and the provision of personalized services to patients. 

% However, the proposed method is applicable to any type of descriptors and could be also
% \begin{figure}[htbp]%figure1
% \centerline{\includegraphics[width=.4\textwidth]{figures/CtsPie.png}}
% \caption{Distribution of descriptors per number of corresponding concepts in MeSH 2018. A considerable amount of descriptors corresponds to multiple concepts.}\label{fig:03}
% \end{figure}
An early version of this work, was presented at the IEEE 32nd International Symposium on Computer-Based Medical Systems (CBMS)~\cite{Nentidis2019_CBMS}. In this extended version, we elaborate on feature generation and selection and present new experiments on balancing and iteratively re-labelling the training dataset. In addition, we investigate the effect of regularisation type in the good predictive performance of logistic regression models and present new results applying the method in a second disease.
The structure of the rest of this article is as follows.
% The rest of this paper is structured as follows. 
In Section \ref{sec:RelatedWork} we provide a brief review of the state of the art in biomedical semantic indexing and a description of the fine-grained semantic indexing  within the landscape of classification under weak supervision.
% In Section \ref{sec:RelatedWork} we briefly review the state of the art in semantic indexing of biomedical literature and describe the fine-grained semantic indexing problem in the context of weakly supervised classification. 
In Section \ref{sec:Methods} we introduce the research questions motivating this work and present the adopted approach.
% In Section \ref{sec:Methods} we present the research questions driving this work and the approach that we have adopted. 
In Section \ref{sec:Experiments} we provide a description of the experimental procedure and a discussion on the corresponding results. 
% In Section \ref{sec:Experiments} we describe the experimental procedure and discuss the corresponding results. 
Lastly, in Section \ref{sec:Conclusion} we draw conclusions on the basis of the results presented. 
% Finally, in Section \ref{sec:Conclusion} we draw conclusions on the basis of the results presented. 

\section{Related work}

% \todo[inline]{Add more about \textbf{Snorkel} team work, potentially reduce details based on Zhou's classification.

% Common

% - Exploit heuristic patterns for automated development of weak supervision (labeling functions)

% - different labels of granularity are considered

% Different

% - We work with a single source of ("explicit") weak supervision (labeling function)

% - We exploit ``coarse-labels'' narrowing the problem in "fine-grained" level, while MeTaL handles the case as a mult-task problem.

% - We focus on the contribution of the weak supervision itself (labeling function) as feature too.

% }

\label{sec:RelatedWork}
% Fine-granularity of semantic annotations at concept level has remain beyond the main focus of recent advancements in the field of semantic indexing of biomedical literature. This is partly due to the lack of adequate data for training and evaluation of supervised machine learning approaches. Below, we briefly review the state of the art in biomedical semantic indexing and investigate the position of fine-grained semantic indexing in the landscape of the weakly supervised classification field.   
In this section we briefly review the state of the art in semantic indexing of biomedical literature and examine the positioning of fine-grained semantic indexing problem within the context of classification under weak supervision.

\subsection{Semantic indexing of biomedical literature}

The main focus of research on semantic indexing of biomedical literature has been on the identification of appropriate MeSH labels for biomedical articles.
% Research on semantic indexing of biomedical literature has mainly focused on identifying the appropriate MeSH entries for each biomedical article. 
This is in accordance with the current practice in NLM, where PubMed/MED\-LINE citations are manually annotated with the appropriate MeSH labels by expert human indexers. 
% This is based on the current practice of NLM, where expert human indexers manually annotate PubMed/MEDLINE citations with the appropriate MeSH entries. 
Currently, the number of available articles with manual annotations exceeds 13 million.
% At the time of writing this paper, there are more than 13 million articles available with corresponding abstract text and manually assigned MeSH entries. 
This is an important resource, suited for the development of machine learning methods that can automatically annotate biomedical articles with MeSH entries, mostly descriptors, driving to solutions of high accuracy.
% This valuable resource has been used for the development of machine learning systems that automatically assign MeSH entries, especially descriptors, to biomedical articles, leading to the development of highly accurate solutions. 
Such systems are necessary for human annotators to keep up with the ever-increasing volume of literature published. 
% Human annotators need such systems to keep up with the ever-growing amount of published literature. 
Provided that this kind of automated solutions are sufficiently good, manual annotation could be reduced significantly.
% In cases where such automation works sufficiently well, the human annotation stage could even be omitted.
% \todo[inline]{stress the motivation: we need systems for helping the annotators, and replacing junior annotators for journals where automation works well} 
For the state of the art performance of such annotation systems see the recent results \cite{Nentidis2019_BioASQ} of the semantic indexing task of the BioASQ challenge \cite{tsatsaronis2015overview}.

% For fine-grained semantic indexing no such dataset exists, as annotations beyond the level of MeSH descriptors detail is beyond the current practice in semantic indexing of scientific articles.
The importance of fine-grained semantic indexing for precision in information retrieval has been highlighted by Darmoni \textit{et al.} \cite{Darmoni2012}.
% Darmoni \textit{et al.} \cite{Darmoni2012} highlighted the importance of fine-grained semantic indexing for precise information retrieval. 
They experimented with chronic and rare diseases, looking only for literal occurrences of corresponding MeSH concept terms in the abstract or title of articles.
% For their experiments on rare and chronic diseases, they assume that the articles that contain literally some term of a MeSH concept in their abstract or title are the only ones that should be indexed with  this concept. 
Despite the strong assumption of literal occurrence, they concluded that indexing at the level of MeSH concepts is beneficial, in terms of precision in the retrieval of relevant documents and incorporated it in the indexing policies for librarians of the CISMeF catalogue\footnote{\url{http://www.chu-rouen.fr/cismef/}} for French medical resources on the Internet. 
% On the other hand, the manual indexing of scientific articles in PubMed/MEDLINE is still done at MeSH descriptor level. Given these MeSH annotations and the concepts corresponding to each descriptor, we aim to exploit article information to automatically identify which of these concepts are actually relevant to the subject of the article, extending the existing annotation in a more fine-grained level of detail. 
% Similar ``dictionary-based'' approaches, based on literal occurrence of label terms and associated tokens, have also been applied in MeSH subheading indexing \cite{Neveol2007}. 

The existing relations of MeSH concepts with corresponding descriptors have also been exploited for the automated semantic indexing of biomedical literature with MeSH descriptors. Specifically, The component of MTI named ``Restrict to MeSH'' \cite{Bodenreider1998} maps concepts automatically extracted from the text of articles to the most relevant MeSH descriptors. Nevertheless, here the final annotations are still at a coarse level.
% for fine-grained semantic indexing we deal with the opposite problem. That is, to extend an existing coarse MeSH subject annotation to more fine-grained concept subject annotations. 

% \textcolor{gray}{[Optional paragraph] }
Similar problems have been studied in the field of fine-grained Named Entity Recognition, e.g. classifying recognized person instances into specific categories~\cite{Fleischman2002} and more generally, inducing the semantic sub-type of extracted noun phrases~\cite{Rahman2010,Ling2012}. Other similar tasks are the ``Type-Compatible Grounding''~\cite{Zhou2017} of unseen entities to similar entities in Wikipedia and
the recently studied ``Entity-Aspect Linking''~\cite{Nanni2018}, where given a textual mention of a named entity, the goal is to identify which section of the corresponding Wikipedia article is relevant to the specific mention. Approaches based on different levels of supervision have been adopted in these tasks, including weakly supervised~\cite{Fleischman2002,Ling2012} and zero-shot learning~\cite{Zhou2017}. 
The similarity of fine-grained semantic indexing with these problems is that given the coarse class of an instance, only valid fine-grained labels are considered. The difference in this case is that classes concern documents, instead of named entities (text spans).

% \todo[inline]{(additional) notes about hierarchical text classification:

% “Single-label Leaf Classification on Tree Hierarchies” (Kosmopoulos 2015)
% \begin{itemize}
%     \item Focus on large scale 
%     \item Cascading : One binary classifier per node vs its siblings 
%     \item Probabilistic Cascading: Considering the whole path root-to-leaf
%     \item Feature selection at top levels can decrease accuracy.
%     \item Use PCA instead of feature selection (for siblings)
% \end{itemize}
% Hierarchical relations between labels has been exploited for Electronic medical record classification into ICD9 categories, especially for infrequent labels (few or zero shot problems)   \cite{Rios2018} ``Few-Shot and Zero-Shot Multi-Label Learning for Structured Label Spaces''.
%     \begin{itemize}
%         \item the ``emerging'' label problem
%     \end{itemize}
%  Concept Labeling: Expert supervision in the form of a mapping between concepts in an ontology to the target classes of interest \cite{Chenthamarakshan2011}
%     \begin{itemize}
%         \item ``even if the ontology-based classifier only weakly approximates the true underlying Bayes optimal classifier, the labels it generates can induce a strong classifier in the bag-of-words represen- tation.''
%     \end{itemize}
% }

\subsection{Classification under weak supervision}

In supervised machine learning, a training dataset with ground truth annotations is used to develop a model for the annotation of unseen instances.
However, in lack of such a dataset, models can also be learned under partial or weak supervision. A variety of weakly supervised settings have been proposed where the weak supervision may derive from different resources such as heuristic rules, expected label distributions, crowd-sourcing and reuse of existing resources \cite{AlexRatnerStephenBachParomaVarma}. 
% However, in lack of such a dataset for fine-grained semantic indexing, models can also be learned from partially or weakly labeled data. A variety of weakly supervised approaches have been studied and three typical cases have been recently described by Zhou\cite{Zhou2018}, namely inaccurate, incomplete and inexact supervision.

The typical case, where only a small part of the dataset is labelled, is often considered as incomplete supervision. In this case, unlabelled data are exploited by semi-supervised learning approaches, to compensate for the lack of training data \cite{Nigam2001}. 
In fine-grained semantic indexing no labeled data are available at all, therefore the incompleteness of the supervision is not the focus of this work.
% In the problem studied here, the abstracts that remain unlabeled after obtaining weak labels are few compared to the number of weakly labeled instances. 
% In the problem studied here, the abstracts that remain unlabeled after obtaining weak supervision, i.e. where none of the concepts corresponding to the MeSH descriptor occurs, are few compared to the number of weakly labeled instances. 
% Therefore, the incompleteness of the supervision is not the focus of the proposed method. 
% which aims at a sufficiently general model learned from inaccurate supervised data. This model will also be able to predict labels for the unlabeled instances.

Additionally, multi-instance learning is often seen as a case of weak supervision \cite{Zhou2018}. In this case the ground truth labels are coarse-grained, in the sense that they are assigned at the level of bags of instances, while the instances in the bag are unlabelled. The latter can be assigned weak labels.
% In inexact supervision, each labeled example, also called a bag, consists of multiple instances. The ground truth example labels are coarse-grained, in the sense that they are assigned at the level of the bag, while the instances of the bag are unlabeled. This is why this multi-instance learning setting is considered by some authors as a case of weak supervision\cite{Zhou2018}. However, no such type of weak supervision arises in the problem studied in this paper.
% In the typical case, the target labeling is also at example level, though detection of key instances inside examples has also been studied\cite{Liu2012}. 
In fine-grained semantic indexing no natural structure exists in grouping the articles into bags, therefore we focus on single-instance approaches. 
% Therefore, a single-instance approach is adopted for the proposed framework and the coarse-grained labels, at the MeSH descriptor level, are only exploited for the construction of the training datasets,

% This is different from candidate labels problems that are usually Single-Label and presence of a label in different sets is usually exploited to produce a common model for all labels\cite{Cour2011}. For fine-grained semantic indexing each fine-grained label is always accompanied with the same set of candidate labels, that is the set of the rest concepts corresponding to the same MeSH descriptor. Therefore, we focus on the development of one model per MeSH descriptor. 

% \todo[inline]{(additional paragraph) Noise effect and classification in this case:}

Of particular interest are methods that treat some of the labeled instances as being potentially erroneous and untrustworthy \cite{Brodley1999}. 
% In other cases, labels can be available for all the training instances, but some of them are erroneous and should be tackled as noise \cite{Brodley1999}.
When using heuristic-based weak labelling for fine-grained semantic indexing, erroneous labels are to be expected. 
% The problem of fine-grained semantic indexing pertains to this category.
% Some of the heuristically assigned fine-grained training labels are expected to be erroneous.
% The fine-grained training labels are assigned heuristically to the instances and a proportion of them is expected to be wrong. 
% In particular, false negative labels are expected when no corresponding concept occurs in the abstract of relevant articles and false positive labels due to occurrence of concepts in the abstract of irrelevant articles.
% In particular, false positive examples are expected because a concept may occur in the abstract of irrelevant articles, and false negative examples due to the lack of concept occurrence in the abstract of relevant articles. 
% Weaknesses of the tools employed for automated concept occurrence recognition are also expected to produce errors that will inflate both false negative and false positive cases.
% Both cases are also expected to be inflated by errors of the concept occurrence recognition tools employed.
% This kind of errors is usually handled as labeling noise and an approach to minimize their effect in supervised learning is to identify and filter out potentially erroneous labels \cite{Brodley1999}. Towards this direction, similarity among instances have been exploited, in terms of geometrical vicinity \cite{Muhlenbach2004}, 
% The proposed method aims at a sufficiently general model learned from these inaccurate supervised data, that will also be able to predict labels for the unlabeled instances.
% \textcolor{gray}{[Optional paragraph]}
The presence of noise in the training datasets, especially label noise, is expected to decrease the predictive performance of the trained classification models. The level of noise for the different features and the classes of the training data is important. 
% Some concept terms are more ambiguous than others, and the heuristic labeling may be more prone to errors in articles containing these terms. As a result, noise distribution may depend both on the class and the attribute values of the instances.
% and can be classified as ``Noise Not at Random'' according to the taxonomy proposed by Frenay and Verleysen \cite{Frenay2014}.}

A range of approaches have been suggested to tackle the negative effects of label noise in classification. 
Some of them use filtering to identify potentially mislabeled examples in the training dataset. This kind of filter is usually based on the labels of close neighbours (similar instances) \cite{Muhlenbach2004} or exploit the disagreements in the prediction of classifiers trained on different portions of the dataset \cite{Brodley1999, Ratner2017}. 
% The idea of using disagreements has also been extended in the combination of noisy supervision from diverse resources, including heuristics based on patterns, crowd-sourcing, or distant supervision \cite{Ratner2017}.
% from external databases 
% or  \cite{Ratner2017} .
% and hierarchical weak supervision at different levels of granularity \cite{Ratner2018}.

In particular, Snorkel MeTaL \cite{Ratner2018} considers hierarchical weak supervision at different levels of granularity, exploiting coarse and fine-grained labels, as required for fine-grained semantic indexing. However, Snorkel MeTaL formulates a multi-task problem unifying the classification at different levels, while in fine-grained semantic indexing we narrow down the problem in the fine-grained level. In addition, while the focus of Snorkel MeTaL is on the combination of multiple different resources of weak labelling, named labelling functions, for fine-grained semantic indexing we have no alternative supervision resources available, therefore we focus on the potential of a single heuristic for weak labeling and investigate the complications of this practice.

% In this work, we exploit the existing coarse-grained labels to narrow down the problem in the fine-grained level, while MeTaL formulates a multi-task problem unifying the classification at different levels of granularity. In addition, while the focus of MeTaL is on the combination of multiple different resources of weak labelling, named labelling functions, we focus on the potential of a single heuristic, as labelling function and feature at the same time, and investigate the complications of this practice.

There are learning algorithms that have been specifically designed to model and withstand noise of specific types, diminishing its effect on predictive performance~\cite{Angluin1988, Natarajan2014}. However, they usually assume the noise to be random, which is not expected to be the case for the weak labels based on concept-term occurrence, as some concepts and terms are more ambiguous than others.
On the other hand, algorithms that were not specifically designed to tolerate noise can also be robust to certain noise types in practice, particularly when 
techniques for overfitting avoidance are used, like bagging in decision-trees with post-pruning~\cite{Abellan2010}.
% \todo[inline]{Alternative example: ... like random forests (ref?)}
% In the problem studied here explicit modeling of noise is complicated by the dependency of noise distribution both on the class and the attribute values of semantic features, as some concept terms are more ambiguous than others.
Based on this idea, in this work we study the use of standard learning algorithms with weak supervision, rather that adopting elaborate noise modeling approaches.
% without dealing with the explicit and elaborate modelling of the noise distribution, which depends both on the class and the attribute values of features, as some concepts are more ambiguous than others.
% Based on this idea, in this work we study the use of standard learning algorithms with weak supervision, without dealing with the explicit and elaborate modelling of the noise distribution, which depends both on the class and the attribute values of features, as some concepts are more ambiguous than others.

% \todo[inline]{(additional) material on bootstrapping/iterative approaches:
% \begin{itemize}
%     \item ``Text classification from unlabeled documents with bootstrapping and feature projection techniques'' \cite{Ko2009}
%     \item ``Improving Distant Supervision for Information Extraction Using Label Propagation Through Lists'' \cite{Bing2015}
% \end{itemize}
% }
%GT checked up to here 14/2/2019

\section{Methods}
\label{sec:Methods}

As golden fine-grained semantic annotations are not available for the development of prediction models, we propose an approach that exploits weak labels, automatically extracted from the article text, as weak supervision. This approach also exploits the available manually assigned MeSH topics and the known internal structure of descriptors and concepts in MeSH. 
% following the supervision taxonomy proposed by Hernandez-Gonzalez
% \textit{et al.}\cite{Hernandez-Gonzalez2016}. 
Each MeSH descriptor is modelled separately and a specific set of fine-grained labels is known beforehand, based on the conceptual structure of the descriptor.

\begin{figure}[!t]%figure1
\centerline{\includegraphics[width=1\textwidth]{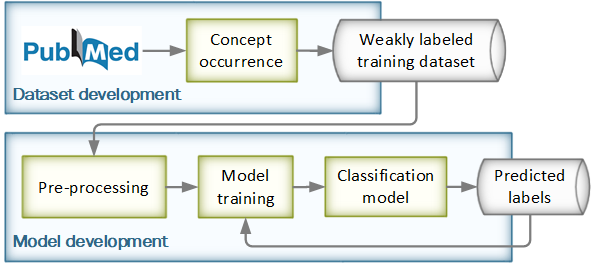}}
\caption{Proposed approach for fine-grained semantic indexing of biomedical articles based on weak supervision.
}\label{fig:block_diagramm}
\end{figure}

The proposed approach, illustrated in Fig.~\ref{fig:block_diagramm}, consists of two phases.
In the phase of dataset development, fine-grained subject annotations are heuristically assigned to articles, based on the occurrence of specific concepts in the text. In the phase of model development, these annotations are used to train predictive models considering lexical and semantic features of article abstracts.
We formulate the problem of assigning fine-grained semantic labels as a single-instance multi-label classification problem, where the weak labels are available at both the stage of training the model and at prediction time.

The specific research questions driving this work are the following:
\begin{itemize}
    \item 
    % Is concept occurrence a good estimation for fine-grained concept-level semantic indexing? 
    Is the occurrence of specific concepts in the articles a competent heuristic for assigning concept-level subject annotations? 
    % To what extend do the heuristically assigned labels coincide with manually-assigned 
    How well do the fine-grained labels assigned by this heuristic approximate the golden labels, as assigned by domain experts?
    \item 
    % Is it possible to exploit concept occurrence as weak supervision to train models for fine-grained semantic indexing without golden training data? 
    Is it feasible to develop models for concept-level semantic indexing using the heuristically assigned labels as weak supervision?
    % Could these models exceed the predictive performance of concept occurrence itself and improve further by training on the predictions?
    Are the predictions of these weakly supervised models better than the weak supervision alone? 
    Could these models improve further by training on their own predictions?
    \item 
    % What feature representations are adequate for weakly-supervised fine-grained semantic indexing models and what is the effect of label imbalance in their performance?      
    What features are useful for modelling fine-grained subject annotations with weak supervision and what is the effect of label imbalance to predictive performance?      
\end{itemize}

% \subsection{A weakly-supervised method for fine-grained semantic indexing} 
\subsection{Dataset development}

% As shown in in Fig.~\ref{fig:block_diagramm} 
% Fig.~\ref{fig:block_diagramm} provides an overview of the proposed approach.
% The proposed approach is illustrated in Fig.~\ref{fig:block_diagramm}.
% First, all articles relevant to a MeSH descriptor \textit{t} are retrieved from PubMed/MEDLINE. This is done through the semantic search functionality provided by Entrez E-Utilities\footnote{\url{https://www.ncbi.nlm.nih.gov/books/NBK25501/}} based on manual MeSH annotations. 
As a fist step towards dataset development, the Entrez E-Utilities\footnote{\url{https://www.ncbi.nlm.nih.gov/books/NBK25501/}} are exploited to  retrieve from PubMed/MEDLINE all the articles that are manually annotated with a specific MeSH descriptor \textit{t} of interest.
% In this work we only consider descriptors, where the preferred concept  \textit{c\textsubscript{pref}} is the broader of all concepts \textit{c\textsubscript{i}} in the set \textit{C\textsubscript{t}} of MeSH concepts corresponding to descriptor \textit{t}. 
% That is, all other concepts in \textit{C\textsubscript{t}} are narrower to \textit{c\textsubscript{pref}}.  
% Then, noisy fine-grained labels are assigned to the selected articles, based on concept occurrence, to develop a weakly-supervised (WS) training dataset. 
Subsequently, the retrieved articles are heuristically annotated with noisy concept labels, employing a concept recognition tool such as MetaMap \cite{Aronson2001}, to develop a weakly-supervised (WS) training dataset. 
% In particular, each article is labeled with all the concepts \textit{c\textsubscript{i}} from \textit{C\textsubscript{t}} that occur in the article. 
Specifically, for each concept \textit{c\textsubscript{i}} from \textit{C\textsubscript{t}} that occurs in an article, a corresponding weak label is assigned to this article.
%  The occurrence  of \textit{c\textsubscript{i}} in  the  article  does  not  guarantee  that  the  article  is  actually relevant  to  this  concept,  even  though  the  article  is  relevant  to  the  descriptor \textit{t}. 

The fact that a concept, \textit{c\textsubscript{i}}, occurs in an article is neither necessary nor sufficient to conclude that this article should be given that concept label.
In particular, false negative labels are expected when no corresponding concept occurs in the abstract of relevant articles and false positive labels due to occurrence of concepts in the abstract of irrelevant articles.
% In particular, false positive examples are expected because a concept may occur in the abstract of irrelevant articles, and false negative examples due to the lack of concept occurrence in the abstract of relevant articles. 
Weaknesses of the tools employed for the automated concept occurrence recognition are also expected to produce errors that will inflate both false negative and false positive cases.
%  However, we expect that the occurrence of \textit{c\textsubscript{i}} will be highly correlated with relevance to \textit{c\textsubscript{i}}, and can be used for noisy fine-grained labeling of articles.
Nevertheless, it is expected that \textit{c\textsubscript{i}} occurrence will be more frequent in articles relevant to \textit{c\textsubscript{i}} and relatively rare to non-relevant articles. Therefore, we consider it as a good heuristic for assigning noisy concept-level labels to articles relevant to \textit{t}. 
% The noise on these labels is expected to depend both on the class and the attribute values of features rather than being random, as some concepts and terms are more ambiguous than others.

% \todo{Idea: Could articles with \textit{c\textsubscript{i}} occurrence but not relevant to \textit{t} (if any) exploited as negative examples?}

% The identification of biomedical concept occurrences in text is an information extraction task, involving the recognition of biomedical named entities and their mapping to specific concepts in a normalized semantic system. 
The information extraction task of identifying the occurrence of biomedical concepts in natural language text involves biomedical named entity recognition and mapping of each recognized entity to a concept.
% Particular challenges of this task include the recognition of concepts with multi-word terms or terms appearing inflected in the text, and disambiguation of terms belonging to more than one homonymous concepts. 
Specific challenges and sources of noise in this process include identification of multi-word terms, term inflection, and term disambiguation in case of homonymous concepts.
% Automated identification of biomedical concept occurrence in text has been in the focus of biomedical natural language processing community and a variety of approaches and tools have been proposed for this task \cite{Jovanovic2017}.
Significant efforts by the community of biomedical natural language processing have led to a range of proposed methods and tools for automatically identifying the occurrence of biomedical concepts \cite{Jovanovic2017}.
% In this work, we employ MetaMap \cite{Aronson2001}, one of the most popular and comprehensive approaches, to recognize occurrence of Unified Medical Language System (UMLS)\footnote{\url{https://www.nlm.nih.gov/research/umls/}} concepts. MeSH concepts are a subset of UMLS concepts and they are linked directly. 
In the context of this work, MetaMap, one of the most popular and comprehensive tools in the field, is employed for concept-occurrence extraction. MetaMap links concept occurrences to concepts of the Unified Medical Language System (UMLS)\footnote{\url{https://www.nlm.nih.gov/research/umls/}}, which are a super-set of MeSH concepts and are also directly linked to them.

% Based on this noisy labeling we propose a framework for the development of ML models that automatically provide refined topic annotations considering information from the abstract and the title of the articles. 
% For this purpose MetaMap is employed, which is a state-of-the-art biomedical information extraction tool developed in NLM. MetaMap recognizes concept occurrence in the articles considering a variety of parameters including concept synonyms, alternative forms, partial matches and disambiguation with homonyms.

% Since each article considered for fine-grained indexing is already indexed with \textit{t} we assume that at least one \textit{c\textsubscript{i}} from \textit{C\textsubscript{t}} is relevant to the article. In particular, each article is at least related to the broader \textit{c\textsubscript{top}} concept.
As the articles considered in the task are manually annotated with \textit{t}, they could all trivially be labelled with \textit{c\textsubscript{top}} which is the broader concept in \textit{C\textsubscript{t}}.
% However, the recognition of this ``default'' class \textit{c\textsubscript{top}} is not useful and it is therefore not considered as one of the fine-grained labels to be predicted. 
Therefore, predicting relevance to this ``default'' most general label is not useful and \textit{c\textsubscript{top}} is not included in the set of fine-grained labels to be predicted. 
% Articles containing occurrences of \textit{c\textsubscript{top}} are included in the dataset but their \textit{c\textsubscript{top}} annotations are ignored for model development and validation.
The datasets include articles with \textit{c\textsubscript{top}} occurrences but no \textit{c\textsubscript{top}} weak labels are assigned to them.
% , neither are considered for the development and testing of models.
For example, considering as \textit{t} the descriptor for ``Alzheimer Disease'' (AD), the \textit{C\textsubscript{t}} consists of AD, which is the \textit{c\textsubscript{top}}, and six narrower concepts such as familial, early onset and late onset AD, which correspond to specific types of AD. In this case, there is no need to predict the \textit{c\textsubscript{top}} AD concept as all the articles are relevant to it.

\subsection{Model development}
% The abstract and the title of each article in the weakly-labeled dataset are exploited to produce two types of features for the articles. 
For all the articles in the weak supervision dataset both lexical and semantic features are produced, using the title and the abstract of the articles.
% Though the full text of some articles is also available in PubMed Central\footnote{\url{https://www.ncbi.nlm.nih.gov/pmc/}}, we currently focus our analysis on title and abstract which are available for more articles and because concepts occurring there are expected to be related to the main topic of the article, in contrast with concepts found in its main body. 
Although for some articles the main body is available as well, through PubMed Central\footnote{\url{https://www.ncbi.nlm.nih.gov/pmc/}}, our analysis focused on titles and abstracts. This is because we expect that concepts found in the title and the abstract of an article are more relevant to its main subject than concepts occurring in its full text. In addition, using only the title and the abstract allows us to uniformly exploit a larger set of articles, for which the full text is not available.

\begin{figure}[!t]%figure2
\centerline{\includegraphics[width=0.8\textwidth]{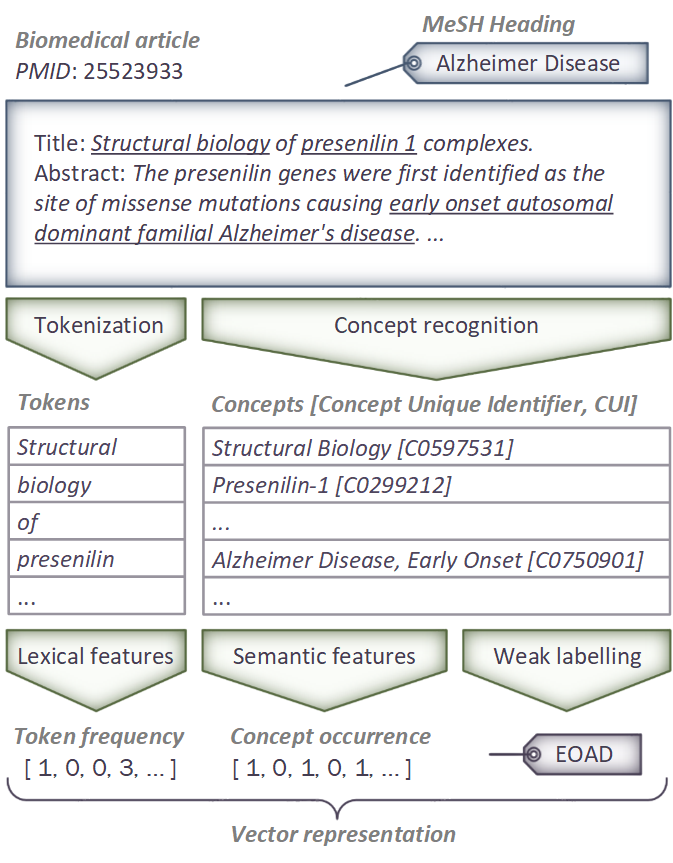}}
\caption{Lexical and semantic feature generation and weak fine-grained labeling for an article relevant to AD based on its title and abstract.
% \todo[inline]{Should I add a reference for this paper?}
}\label{fig:feature_extraction}

\end{figure}
% In particular, as presented in Fig.~\ref{fig:feature_extraction} the text is tokenized and the number of occurrences is calculated for each token, producing bag-of-words lexical features. 
An example of feature generation for an article is presented in Fig.~\ref{fig:feature_extraction}.
Lexical features are produced by tokenizing the text and counting the number of token occurrences, in a bag-of-words approach.
% In addition, the occurrence of concepts in the text, extracted with the use of MetaMap\footnote{Local installation of MetaMap2016 V2 called through SemRep V1.7 with the 2015AA UMLS vocabulary resources.}, provides additional semantic features.
Concept occurrences, on the other hand, are extracted from the text using MetaMap\footnote{MetaMap2016 V2 (through SemRep V1.7) using 2015AA UMLS Metathesaurus.} and they constitute the semantic features of the articles.
In particular, semantic features are produced for any occurring UMLS concept, regardless of its relevance to the topic \textit{t} under study. 
% These features are binary, indicating whether a concept occurs in the abstract rather than counting the number of occurrences.
% Preliminary experiments with absolute concept-occurrence count yielded lower performance.

% All extracted UMLS concepts are considered as features, regardless of their resource vocabulary or semantic type, not only the ones corresponding to the MeSH descriptor of interest. 
Both semantic and lexical features are weighted based on their scarcity in the WS dataset applying TF-IDF transformation.
% On both lexical and semantic features TF-IDF transformation is applied to weigh the features, based on their scarcity in the dataset. 
For concept-occurrence features, the binary frequency is used for the TF-IDF transformation, which has value one if the concept occurs at least once and zero otherwise, indicating whether a concept is present or not. Preliminary experiments with absolute concept-occurrence count yielded lower performance.
% For concept-occurrence features which are binary, the Boolean term frequency is used for the TF-IDF transformation.
% The resulting vector representation of each article consists of one feature for each distinct token found in any of the articles and one for each distinct concept recognized in any of them. An upper bound for concept features is the size of MetaMap vocabularies which consists of more than three million concepts.
% Some of these features are not informative enough to be useful for model development and some of them may also introduce noise.
% As some features may be less informative or may introduce noise we use feature selection to select the most useful ones, based on their ability to discriminate between the target classes in the training data.
In addition, feature selection is used to keep only the most informative and useful features and disregard the ones that do not help to discriminate between the target labels in the WS dataset and may introduce noise.
% For this reason, we use feature selection to select the most useful ones, based on their ability to discriminate between the target classes in the training data. 
% In particular, for an estimation of this ability we experimented with three alternative statistics for dependency between each feature and the target classes. Namely, the Chi-squared statistic, the ANOVA F-measure and the Mutual Information. In addition, we also experimented with different numbers of selected features ranging from 5 to 1000 features. For each combination of feature selection method and number of features, a different version of the training dataset was developed and a corresponding version of the validation dataset, with the same set of selected features.
% Only these selected features are used to produce the final vector representation of the articles, which is used for model development.
% The final vector representation of each article is produced based on these selected features only and it is used for the development of classification models.

% Since the task of fine-grained semantic indexing is multi-label, we adopt a One-Versus-Rest approach, where a distinct binary classifier is trained for each of the labels corresponding to each concept from \textit{C\textsubscript{t}} excluding \textit{c\textsubscript{top}}. 
Adopting a One-Versus-Rest approach for the multi-label task of fine-grained semantic indexing, we train a distinct binary classifier for each \textit{c\textsubscript{i}} label, excluding \textit{c\textsubscript{top}}.  
% Articles annotated with \textit{c\textsubscript{top}} in the weakly-labeled dataset are kept in the dataset but their \textit{c\textsubscript{top}} annotations are ignored.
Therefore, each article is assigned (or not) a fine-grained subject label by the corresponding label-specific binary model. All the predictions are combined to produce a multi-label annotation for the article.
% At the prediction phase, each class-specific model predicts the relevance of an article for the corresponding fine-grained label and the predictions are integrated to produce a final set of all predicted fine-grained subject labels for each article. 
Finally, we also investigate the iterative training of new predictive models using the predicted fine-grained subject labels in the place of the initial WS labels.
% Finally, we also investigate some configurations where the predicted fine-grained labels are then used in the place of the initial weak labeling to train new classification models in an iterative approach. 

% This procedure of model development is summarized in Fig.~\ref{fig:07}.
% \begin{figure}[!h]%figure7
% \centerline{\includegraphics[width=.46\textwidth]{figures/Schema_ModelDevelopmentFlat.png}}
% \caption{The stages of model development}\label{fig:07}
% \end{figure}

\section{Experiments}
\label{sec:Experiments}
The proposed method has been implemented in Python, using the scikit-learn\footnote{\url{https://scikit-learn.org/}} library, and a series of experiments has been carried out. 
% The proposed method has been implemented in Python, using the SciKitLearn\footnote{\url{https://scikit-learn.org/}} library, and has been used to carry out the experiments presented below.
This implementation is  openly available\footnote{\url{https://github.com/tasosnent/BeyondMeSH}}, as well as all the datasets and model configurations for the experiments which are reported below. 
% The implementation, the model configurations and the datasets for the reported experiments are openly available\footnote{\url{https://github.com/tasosnent/BeyondMeSH}}. 
\subsection{Experimental setup}
\label{ssec:models}
The proposed method was applied independently to two different diseases, the Alzheimer's Disease (AD) and the Duchenne Muscular Dystrophy (DMD). The experimental setup, which is similar for the two use cases, is described in the respective subsections. We are interested to identify and justify differences in the behaviour of the method in the two use cases.
% In particular, regarding the second use-case we focus on stressing the differences with the first use-case.

\subsubsection{Experiments on Alzheimer's Disease}
\label{sssec:ADmodels}

The first application of the proposed method has been for the MeSH descriptor ``Alzheimer Disease'' (AD). 
The \textit{C\textsubscript{t}} in this use case, consists of the homonymous AD concept and six narrower ones, namely, Presenile Dementia (PD), Focal-onset AD (FOAD), Early-onset AD (EOAD), Late-onset AD (LOAD), Familial AD (FAD) and Acute Confusional Senile Dementia (ACSD). The \textit{c\textsubscript{top}} is the same as the \textit{c\textsubscript{pref}}, that is the AD concept. 
% In this case, \textit{C\textsubscript{t}} consists of the homonymous concept, which is both the \textit{c\textsubscript{pref}} and the \textit{c\textsubscript{top}}, and six narrower concepts, namely, Early-onset AD (EOAD), Late-onset AD (LOAD), Focal-onset AD (FOAD), Familial AD (FAD), Presenile Dementia (PD) and Acute Confusional Senile Dementia (ACSD). 

In the experiments, a dataset was gathered comprising 68,542 articles annotated with the AD descriptor\footnote{On 17 Apr 2018, searching with MeSH descriptor id D000544} from PubMed, as well as  their abstract and title.
% Weak labels have been assigned to 51,450 of them, leaving 17,092 articles without labels.
The heuristic weak labeling procedure, based on the \textit{c\textsubscript{i}} occurrence, assigned weak fine-grained labels to 51,450 of the articles and left the rest 17,092 articles without any fine-grained label. 
The second column of Table~\ref{tabADinitial} summarizes the distribution of articles in the different weak labels.
% The distribution of the heuristic (WS) labels in the initial dataset is summarized in Table~\ref{tabADinitial}. 
No occurrence of ACSD and FOAD was automatically recognized in the title or abstract of any article in the dataset, and we therefore excluded these two extremely rare labels from the experiments.
% In particular,FOAD and ACSD have not been recognized in any of the articles and as a result these two extremely scarce concepts where excluded from model training and validation.
% This specific experiment aims at automated classification of articles, already annotated with the AD descriptor, as actually relevant to any of the specific disease types corresponding to the narrower concepts, excluding the \textit{c\textsubscript{top}} weak labels. 
% The initial goal of the specific experiment is to classify articles annotated with the AD descriptor as actually relevant to any of the narrower disease types, ignoring the \textit{c\textsubscript{top}} weak labels. 
Therefore, we use four fine-grained classes in this use case: PD, LOAD, EOAD and FAD.
% The specific classes in this case correspond to four types of the disease, namely, PD, LOAD, EOAD and FAD.
% These narrower classes are the four disease types: PD, FAD, EOAD and LOAD.

\begin{table}[!t]
\caption{Number of articles per WS label in the AD datasets.}
\begin{center}
\begin{tabular}{|c|c|c|c|c|c|}
\hline
\textbf{Label}&\textbf{Initial}& \multicolumn{2}{|c|}{\textbf{\textit{Test datasets }}}& \multicolumn{2}{|c|}{\textbf{\textit{Training datasets}}} \\
\textbf{annotation}&\textbf{dataset}&\textbf{\textit{ MA1}}& \textbf{\textit{MA2 }}&\textbf{WS} & \textbf{WS\textsubscript{und}}  \\
\hline
AD$^{\mathrm{*}}$   	&   50,233   &  73      & 49        & 50,111  & 3000 \\
PD  	                &   154     &  1         & 18       & 135     & 135\\
FAD                     &   934     &  3       & 33        & 898     & 898\\
EOAD                    &   671     &  0        & 42       & 629     & 629\\
LOAD                    &   371     &  0        & 29       & 342     & 342\\
FOAD$^{\mathrm{*}}$     &   0       &  0        & 0          & 0     & 0\\
ACSD$^{\mathrm{*}}$     &   0       &  0        & 0          & 0     & 0\\
% \hline
labeled 	                    &   51,450   &  75      & 93      & 51,282   &  4170\\
no labels	                    &   17,092   &  25      & 7        & 0       & 0\\
\hline
total        		            &   68,542   & 100      & 100     & 51,282  & 4171 \\
\hline
% \multicolumn{5}{1}{$^{\mathrm{*}}$Labels ignored for model development and testing.}
\end{tabular}

\label{tabADinitial}
\end{center}
\small{$^{\mathrm{*}}$Labels ignored for model development and testing.}\\
\end{table}

% In order to measure classification performance, some ground truth annotations were needed. For this purpose, a random subset of 100 articles (MA1) has been held out of the initial dataset for manual annotation.
Some golden annotations were needed for measuring the classification performance of the weakly supervised models. To this end, a subset (MA1 AD) of 100 randomly selected articles has been left out of the initial dataset to be annotated manually.
Additionally, as the distribution of the weak labels indicated that the initial dataset is highly imbalanced, with most of the articles being labeled with \textit{c\textsubscript{pref}},  
% However, the weak label distribution of the initial dataset suggested that the distribution of classes is strongly skewed, with the majority of the articles labeled with \textit{c\textsubscript{pref}}. 
the classes of interest were expected to be under-represented in the random MA1 AD dataset.
% To handle the expected under-representation of low prevalence classes in the random subset, a balanced subset of 100 articles (MA2) has also been selected, based on the weak labels. 
For this purpose, another dataset (MA2 AD) of 100 articles has been left out, with balanced representation of the weak labels.   

The articles for the balanced MA2 AD dataset have been selected with an iterative procedure considering label combinations (subsets of the set of all available labels) in the initial weakly labeled dataset.
% The MA2 dataset has been built using an iterative procedure based on label combinations, which are the subsets of the set of all available labels.
In this procedure, all the articles where grouped based on the unique combination of their weak labels. Then, for each label combination one of the corresponding articles was selected for inclusion in the MA2 AD dataset. This step was repeated until a total of 100 articles was selected. During these repetitions, if half of the initial articles for a label combination had already been selected for inclusion in MA2 AD, no further articles were selected from this label combination. 
% During this procedure, one article annotated with each label combination was added in MA2 until 100 articles were selected or half of the articles annotated with this label combination were selected. 
The over-represented label combination of \textit{c\textsubscript{pref}} (AD) alone was omitted during the selection procedure. Table~\ref{tabADinitial} presents the distribution of the heuristic (WS) labels on both datasets (MA1 AD and MA2 AD), which were left out for manual annotation and testing. 

Following the removal of MA1 AD and MA2 AD articles from the initial dataset, the remaining 51,282 weakly labeled articles (WS training dataset) were used to train the predictive models for fine-grained semantic indexing of AD articles.
% The remaining 51,282 articles from the initial dataset were used as the WS training dataset for the development of a multi-label classification model to predict concept-level labels for articles relevant to \textit{t}. 
A second version of the WS training dataset (WS\textsubscript{und}), was also developed by under-sampling the set of articles annotated with \textit{c\textsubscript{pref}} (AD) only, reducing both the over-representation of the \textit{c\textsubscript{pref}} label and the size of the dataset. 

Alternative configurations have been considered, with and without feature selection, for training different classification models on the training datasets. 
% Different classification models were trained on the training dataset, considering alternative configurations with and without feature selection.
As regards feature selection the top $k$ features were selected, with $k$ ranging from 5 to 1000, based on the ranking by either the Chi squared (\textchi\textsuperscript{2}) or the ANOVA F statistics. 
% Regarding feature types, either only lexical features or lexical and semantic features together were considered. 
Regarding the types of features considered, models were developed based either on lexical features only, or a combination of lexical and semantic features.
% In addition, models with lexical and semantic features excluding the \textit{c\textsubscript{i}} occurrence, used for the weak labeling, were also developed.
In particular, for each of the alternative configurations regarding the type and number of features and the feature ranking statistic, four distinct models were trained. Namely, a Decision Tree Classifier (DTC), a Random Forest Classifier (RFC), a Linear Support Vector Classifier (LSVC) and a Logistic Regression Classifier (LRC). 
% Furthermore, experiments were also carried out on training some additional models on the articles of the training dataset but with labels predicted by the best models on the MA datasets.
% , with SciKitLearn\footnote{https://scikit-learn.org/}.

\begin{table}[!t]
\caption{Top 30 lexical (L) \& semantic (S) features with ranking (R) by F ANOVA.}
\begin{center}
\begin{tabular}{|l l|l l|l l|}
\hline
\textbf{(R)}&\textbf{Feature}& \textbf{(R)}&\textbf{Feature}& \textbf{(R)}&\textbf{Feature} \\
\hline
1   &   (S) PD                    &  11   & (L) ``onset''     &  21   & (L) ``ps1''     \\
2   &   (S) EOAD                  &  12   & (S)  Mutation     &  22   & (L) ``ps2''     \\
3   &   (S) LOAD                  &  13   & (L) ``mutations'' &  23   & (S) Mutant    \\
4   &   (S) FAD                   &  14   & (L) ``eoad''      &  24   & (S) Presenilins    \\
5   &   (L) ``familial''          &  15   & (L) ``load''      &  25   & (L) ``mutant''  \\
6   &   (L) ``presenile''         &  16   & (L) ``presenilin''&  26   & (L) ``psen1''   \\
7   &   (L) ``fad''               &  17   & (L) ``mutation''  &  27   & (S) Load$^{\mathrm{**}}$     \\
8   &   (S) FAD$^{\mathrm{*}}$    &  18   & (S)  PSEN1 gene   &  28   & (L) ``pdat''    \\
9   &   (S) AD                    &  19   & (S)  Familial     &  29   & (L) ``missense''\\
10  &   (L) ``late''              &  20   & (L)  ``early''    &  30   & (L)  ``ps''     \\
\hline 
% \multicolumn{6}{1}{$^{\mathrm{*}}$UMLS concept C3247466 for the ``FAD'' substance.}\\
% \multicolumn{6}{1}{$^{\mathrm{**}}$UMLS concept C1708715 for the ``Loading Technique
% ''.}
\end{tabular}

\label{tabADfeatures}
\end{center}
\small{$^{\mathrm{*}}$UMLS concept C3247466 for the ``FAD'' substance.}\\
\small{$^{\mathrm{**}}$UMLS concept C1708715 for the ``Loading Technique''.}
\end{table}

Table~\ref{tabADfeatures} presents the top 30 lexical and semantic features selected from the WS training dataset based on the ANOVA F. As expected, the top four selected features are the semantic ones corresponding to the \textit{c\textsubscript{i}} that were used for the weak labeling. As the availability of these features can prevent the models from learning something more than trusting these features, 
additional experiments considering lexical and semantic features, apart from the ones corresponding to any \textit{c\textsubscript{i}} from \textit{C\textsubscript{t}} were performed. 

Some of the selected features are synonymous terms of the corresponding  \textit{c\textsubscript{i}} concepts,
% at least in the context of the specific corpus. For example, 
e.g. the abbreviations ``eoad'' and ``load'' for EOAD and LOAD respectively. Other selected features, like the concept of ``Mutation'' and ``PSEN1 gene'', may represent meaningful associations to specific labels, that capture domain knowledge. Finally, it is also interesting that feature selection can also reveal concept recognition errors. In particular, the ``FAD'' concept selected eighth in the list, corresponds to a chemical named ``FAD'', instead of the Familial AD concept. It seems that, in this corpus, the presence of the chemical ``FAD'' concept in an article, even if it is erroneous, it is useful for our task.
% is useful for identifying noisy FAD weak labels.
% probably a wrongly disambiguated occurrence of the FAD disease-type.
Similarly, the concept ``Load'' for ``loading technique'', selected 27th in the list, is miss-recognised in articles where the ambiguous term ``load'' occurs.
% helps disambiguating the LOAD fine-grained labels.
% high ranking of the concept for the procedure of ``loading'' indicates that the concept is miss-recognised in articles where the ambiguous term ``load'' occurs.

Two experts on Alzheimer's reviewed the 200 articles of the two MA AD datasets and manually assigned fine-grained labels to each of them. 
The two experts had 68 disagreements in 52 articles in total. The macro-averaged Kappa statistic over the four classes of interest, was 0.76.
% The inter-annotator agreement between the two experts was 0.76, using the macro-averaged Kappa statistic over the four classes of interest, with 68 different labels assigned to 52 out of 200 articles in total, for the four labels of interest.
The experts resolved their disagreements together and the consensus annotations were used as the final ground truth fine-grained labels in the MA1 AD and MA2 AD datasets for testing.
Table~\ref{tabADMA} presents the distribution of weak (WS) and consensus manual (MA) fine-grained labels in the articles of the MA AD test datasets, as rows and columns respectively. For example, the lower half of the PD column shows that 19 articles in MA2 were manually labeled with PD in total, while the weak labeling assigned the PD, FAD and EOAD labels to 17, 4 and 5 of them respectively.

\begin{table}[!t]
\caption{Number of articles per weak (WS) and manual (MA) label in the randomly selected MA1 AD and label-set balanced MA2 AD datasets. Manual labelling with the broader \textit{c\textsubscript{top}} concept (AD) is not useful as all articles are related to it. }
\begin{center}
\begin{tabular}{|c|c|c|c|c|c|c|c|}
\hline
\multicolumn{3}{|c|}{}&\multicolumn{4}{|c|}{\textbf{\textit{MA labels}}}&\textbf{\textit{total$^{\mathrm{*}}$ }}\\
\cline{4-7}
\multicolumn{3}{|c|}{}&\textit{ \textbf{PD}}& \textit{\textbf{FAD}}& \textit{\textbf{EOAD}} &\textit{\textbf{LOAD}}& \textit{\textbf{WS labels}} \\
\hline
\multirow{5}{*}{\begin{turn}{90} \textbf{\textit{MA1 AD}} \end{turn}}
&\multirow{4}{*}{\begin{turn}{90} \textbf{\textit{WS labels}} \end{turn}}
&\textit{\textbf{PD}}  	            &  1 & 0  & 1   & 1    & 1     \\
&&\textit{\textbf{FAD}}                 &  0 & 3  & 2   & 1    & 3     \\
&&\textit{\textbf{EOAD}}                &  0 & 0  & 0   & 0    & 0     \\
&&\textit{\textbf{LOAD}}                &  0 & 0  & 0   & 0    & 0     \\
% \hline 
&&\textbf{\textit{no labels}}	                &  0 & 3  & 0   & 1    & 25     \\\cline{2-8}
&\multicolumn{2}{|c|}{\textbf{\textit{total$^{\mathrm{*}}$ MA labels }}}    & 1  & 13 & 5   & 5    & \textit{\textbf{MA1 size:}} 100     \\
\hline

\multirow{5}{*}{\begin{turn}{90} \textbf{\textit{MA2 AD}} \end{turn}}
&\multirow{4}{*}{\begin{turn}{90} \textbf{\textit{WS labels}} \end{turn}}
&\textit{\textbf{PD}}  	            &  17& 7  & 6   & 0    & 18    \\
&&\textit{\textbf{FAD}}                 &  4 & 32 & 21  & 4    & 33    \\
&&\textit{\textbf{EOAD}}                &  5 & 32 & 37  & 15  & 42    \\
&&\textit{\textbf{LOAD}}                &  0 & 14 & 15  & 28   & 29    \\
% \hline 
&&\textbf{\textit{no labels}}	                &  0 & 1  & 0   & 0    & 7     \\
\cline{2-8}
&\multicolumn{2}{|c|}{\textbf{\textit{total$^{\mathrm{*}}$ MA labels}}}  & 19 & 58 & 48  & 30   & \textit{\textbf{MA2 size:}} 100      \\
\hline
% \multicolumn{8}{1}{ 
% \begin{flushleft}$^{\mathrm{*}}$As the task is multi-label, the total may be grater than the sum of a column\end{flushleft}
% }\\
% \multicolumn{8}{1}{
% \begin{flushleft} or row.\end{flushleft}
% }\\
% \multicolumn{9}{1}{l}\\
\end{tabular}

\label{tabADMA}
\end{center}
\small{$^{\mathrm{*}}$As the task is multi-label, the total may be grater than the sum of a column or row.}
\end{table}

% The focus of our framework is towards fine-grained classification for all classes considered, regardless of their prevalence. 
The proposed method treats all fine-grained labels equally, regardless of their prevalence.
Consequently, we opted for the label-based macro-averaged F1-measure \cite{Tsoumakas2009} as an overall performance measure.
% , which weights equally the predictive performance for all labels. 
% In addition to the trained models we validated some simple baseline approaches for comparison. 
For comparison, we also calculated the predictive performance of some simple baseline approaches.
Specifically, two trivial baselines considered are assigning all available labels to all articles (AllAll) and randomly deciding to add a label or not (Random).
% In particular, a trivial baseline is labeling all articles with all available labels (AllAll) and another randomly selecting to add a label or not (Random). 
A more reasonable baseline approach is to just trust the weak labels, based on \textit{c\textsubscript{i}} occurrence (WSLabels), which can leave some articles unlabeled. An extension of this approach is to assign to the remaining unlabeled articles all the available labels (WSRestAll). 
% A stronger approach is to trust the initial weak labels (WSLabels). 
% The latter only produces labels for articles with the respective concept occurrences. As a third approach, we extended the latter by assigning to the unlabeled articles all available labels (WSRestAll). 
% \todo[inline]{add baselines? 10-fold on strong labels, Random-trivial }

Although most prior works in biomedical semantic indexing are not directly applicable in this case, due to the lack of labelled training data, simple forms of some ``dictionary-based'' approaches can still be applied. Such approaches rely on literal occurrence of label terms and associated tokens, which are usually refined and enriched based on labelled corpora, as done in MeSH Subheading Attachment \cite{Neveol2007}. Here, we implemented two simple dictionary-based approaches exploiting the available MeSH concept terms, without any statistical processing. 
These approaches assign a concept-label to an article if any dictionary element associated with the concept literally occurs in the title or abstract of the article.
As dictionary elements in the first approach (DTerms) we use the MeSH concept terms (e.g. ``Presenile Alzheimer Dementia'', etc) and in the second (DTokens) the corresponding tokens (e.g. ``Presenile'', ``Alzheimer'', ``Dementia'', etc). Therefore, we expect that DTerms should be more precise but DTokens should have better recall.
% The first approach (DTerms) assigns a concept-label to an article if any term of a concept literally occurs in the title or abstract  (e.g. ``Presenile Alzheimer Dementia'', etc). Similarly, the other approach (DTokens) assigns a concept-label if any of the corresponding tokens occurs (e.g. ``Presenile'', ``Alzheimer'', ``Dementia'', etc). 

\subsubsection{Experiments on Duchenne Muscular Dystrophy}
The same method was also applied to the second use case, for the MeSH descriptor ``Muscular Dystrophy, Duchenne''. In this case, \textit{C\textsubscript{t}} consists of three concepts. The homonymous concept  (DMD), which is the \textit{c\textsubscript{pref}}, the related concept ``Becker Muscular Dystrophy'' (BMD), and the broader concept ``Duchenne and Becker Muscular Dystrophy'' (DBMD) which is the \textit{c\textsubscript{top}}. DMD and BMD are two rare genetic diseases caused by mutations in the same gene. Although related, the two diseases are distinct, with BMD being in general milder than DMD. 
Therefore, distinguishing the two diseases in indexing and retrieval of publications is of clinical relevance.

In this use case 3,619 articles have been retrieved from PubMed for the DMD descriptor\footnote{On 17 Apr 2018, searching with MeSH descriptor id D020388} and weak labels have been assigned to 3,151 of them as shown in Table~\ref{tabDMDinitial}. Again, the \textit{c\textsubscript{top}} labels are not of interest, and therefore the task consists of classifying the articles as relevant to any of the two diseases: DMD and BMD. A randomly selected MA1 DMD dataset and a ``label-set balanced'' MA2 DMD dataset have been selected using the same procedure as for AD and the remaining labelled articles where used for the development of the WS and the WS\textsubscript{und}  DMD training datasets.  

\begin{table}[!t]
\caption{Number of articles per WS label in the DMD datasets.}
\begin{center}
\begin{tabular}{|c|c|c|c|c|c|}
\hline
\textbf{Label}&\textbf{Initial}& \multicolumn{2}{|c|}{\textbf{\textit{Test datasets }}}& \multicolumn{2}{|c|}{\textbf{\textit{Training datasets}}} \\
\textbf{annotation}&\textbf{dataset}&\textbf{\textit{ MA1}}& \textbf{\textit{MA2 }}&\textbf{WS} & \textbf{WS\textsubscript{und}}  \\
\hline
DBMD$^{\mathrm{*}}$   	&   72   &  3      & 25        & 44  & 44 \\
DMD  	                &   2,813     &  74         & 26       & 2713     & 1000\\
BMD                     &   495     &  16       & 50        & 429     & 429\\
% \hline
labeled 	                    &   3,151   &  86      & 75      & 2,990   &  1,277\\
no labels	                    &   468   &  14      & 25        & 0       & 0\\
\hline
total        		            &   3,619   & 100      & 100     & 2,990   &  1,277 \\
\hline
% \multicolumn{5}{1}{$^{\mathrm{*}}$Labels ignored for model development and testing.}
\end{tabular}

\label{tabDMDinitial}
\end{center}
\small{$^{\mathrm{*}}$Labels ignored for model development and testing.}
\end{table}

The 200 abstracts and titles of the MA1 and MA2 datasets where reviewed by a domain expert and the distribution of manually assigned labels compared to the WS labels is shown in Table~\ref{tabDMDMA}. For inter-annotator agreement estimation, the articles have also been reviewed by a student familiarised with the relevant literature. In particular, the student assigned 78 different labels to 49 out of 200 articles in total for the two labels of interest. The macro-averaged Kappa statistic over the two classes of interest was 0.55, which is lower than the 0.76 observed in the AD datasets. 
This is to some extent reasonable since the Kappa statistic penalises random agreement which is higher in the DMD MA datasets, where the annotations are more balanced.
For performance testing of the models, the annotations of the expert were used.

\begin{table}[!t]
\caption{Number of articles per weak (WS) and manual (MA) label in the randomly selected MA1 DMD and label-set balanced MA2 DMD datasets. Manual labelling with the broader \textit{c\textsubscript{top}} concept (DBMD) is not useful as all articles are related to it. }
\begin{center}
\begin{tabular}{|c|c|c|c|c|c|}
\hline
\multicolumn{3}{|c|}{}&\multicolumn{2}{|c|}{\textbf{\textit{ MA labels}}}&\textbf{\textit{total$^{\mathrm{*}}$}}\\
\cline{4-5}
\multicolumn{3}{|c|}{}&\textit{ \textbf{DMD}}& \textit{\textbf{BMD}} &\textit{\textbf{WS labels}}\\
\hline
\multirow{5}{*}{\begin{turn}{90}  \textbf{  \textit{MA1 DMD}}\end{turn}}
&\multirow{3}{*}{\begin{turn}{90} \textbf{\textit{WS labels }} \end{turn}}
&\textit{\textbf{DMD}}  	                &  63 & 12  & 74      \\
&&\textit{\textbf{BMD}}                    &  14  & 3  & 16      \\
% \hline 
&&\textbf{\textit{no labels}}	                    &  14 & 2  & 14       \\
\cline{2-6}
&\multicolumn{2}{|c|}{\textbf{\textit{total$^{\mathrm{*}}$ MA labels }}}   & 88  & 16    &  \textit{\textbf{MA1 size:}} 100       \\
\hline

\multirow{5}{*}{\begin{turn}{90} \textbf{  \textit{MA2 DMD}} \end{turn}}
&\multirow{3}{*}{\begin{turn}{90} \textbf{\textit{WS labels }} \end{turn}}
&\textit{\textbf{DMD}}  	                &  21& 16  & 26      \\
&&\textit{\textbf{BMD}}                    &  45 & 31 & 50     \\
% \hline 
&&\textbf{\textit{no labels}}	                    &  15 & 13  & 25      \\
\cline{2-6}
&\multicolumn{2}{|c|}{\textbf{\textit{total$^{\mathrm{*}}$ MA labels }}} & 81  & 64   &  \textit{\textbf{MA2 size:}} 100     \\
\hline
% \multicolumn{6}{1}{
% \begin{flushleft}$^{\mathrm{*}}$ As the task is multi-label, the total may be grater than the\end{flushleft}
% }\\
% \multicolumn{6}{1}{
% \begin{flushleft} sum of a column/row.\end{flushleft}
% }\\
% \multicolumn{9}{1}{l}\\
\end{tabular}

\label{tabDMDMA}
\end{center}
\small{$^{\mathrm{*}}$As the task is multi-label, the total may be grater than the sum of a column/row.}
\end{table}

As with the AD use case, different classification models were trained, considering alternative configurations regarding feature selection, feature types, and classification models. The performance of the models was again assessed using the F1-measure, macro-averaged over the two labels of interest and the same baseline approaches were also assessed for comparison. In this case, as the \textit{c\textsubscript{pref}} (DMD) is different from the \textit{c\textsubscript{top}} DBMD two additional baseline approaches were used, exploiting the knowledge that DMD is the preferred label which is expected to be assigned to the majority of the articles.
The first one (AllM) is a trivial approach of assigning only the \textit{c\textsubscript{pref}} (DMD) to all the articles. 
The second one (WSRestM), which is similar to the WSRestAll, 
% which apart from the WSLabels also exploits this knowledge of the major class. 
% In particular, this new approach (WSRestM) 
trusts the initial weak labels (WSLabels), 
% as does WSRestAll,
but assigns to the unlabeled articles the \textit{c\textsubscript{pref}} label (DMD).
% ,exploiting the knowledge that this is the preferred label which is epxected to be assigned to the majority of the articles.

\subsection{Results}
\label{ssec:results}
The results of the experiments for both use cases are presented in this section. 
% As done in experimental setup, for the DMD use case we focus on differences with corresponding results for the AD use case.

\subsubsection{Results on Alzheimer's Disease}

\begin{figure}[!tp]%figure8
% \centerline{\includegraphics[width=.48\textwidth]{figures/MA1&MA2Best.png}
\centerline{\includegraphics[width=1\textwidth]{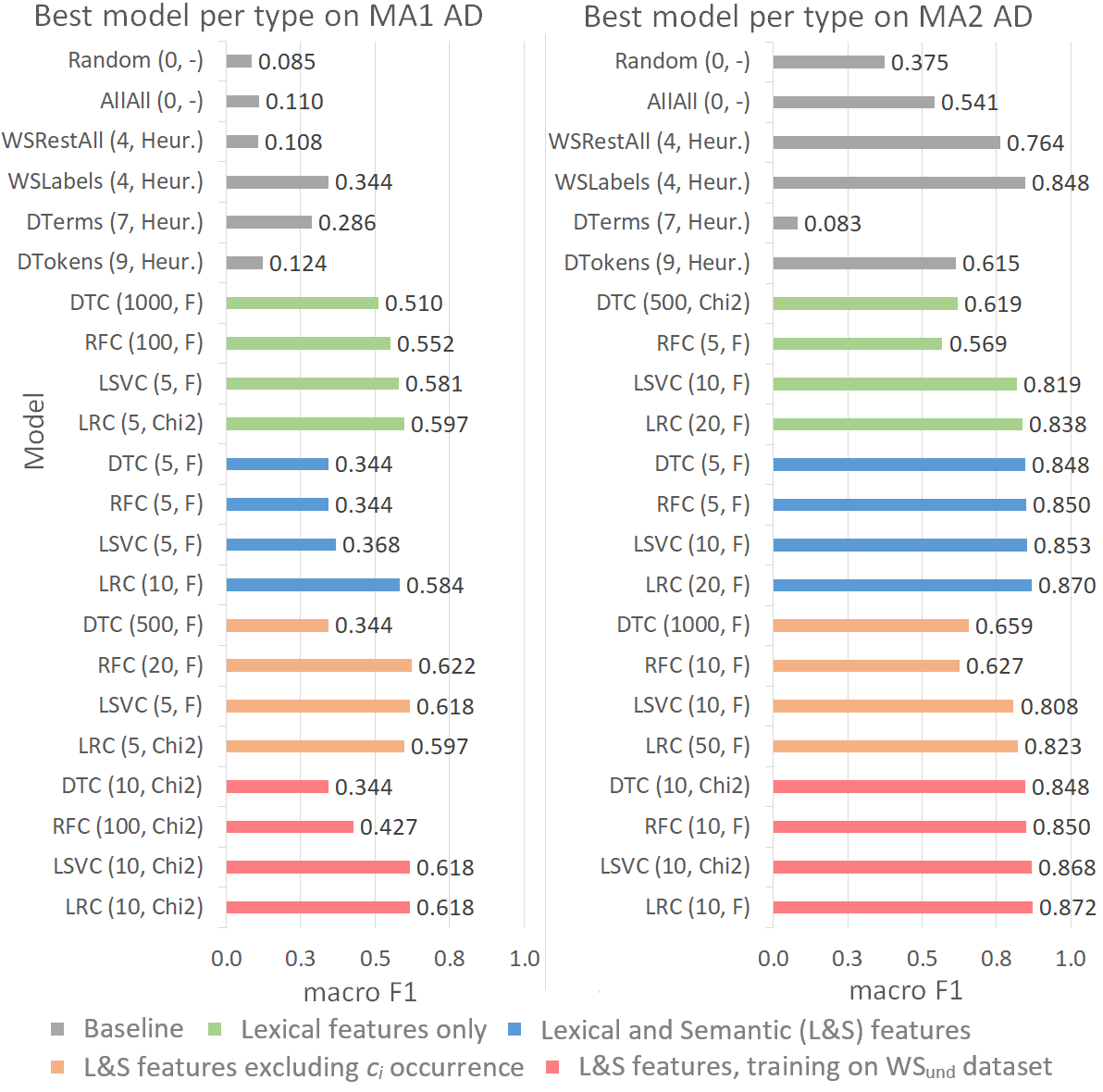}
}
\caption{The performance of the baselines (gray) and the best model per classifier type assessed on the randomly selected MA1 AD dataset (left) and the balanced MA2 AD dataset (right). For each classifier type, models are trained on the WS dataset using lexical features only (green), both lexical and semantic (L\&S) features (blue) and lexical and semantic features excluding the semantic ones that correspond to c\textsubscript{i} occurrence (orange). In addition models using both lexical and semantic features have also been trained in the WS\textsubscript{und} dataset (red). 
Each model is named by the type of the classifier and inside a parenthesis, the number of selected features and the feature selection method separated by a comma. Heur. stands for heuristic selection. 
The F1 measure is macro-averaged over the four labels considered (PD, FAD, EOAD and LOAD).
% The Chi2 and F suffixes in classifier names indicate the feature selection method employed for each model.
}\label{fig:AD_full}
\end{figure}

% Fig.~\ref{fig:MA1&MA2Best} presents the F1 scores of the best model per classifier type mentioned above.
% Fig.~\ref{fig:AD_full} presents the classification performance of the baselines (grey) and the best model per classifier type for each of the experiments described in Section \ref{ssec:models} for the AD use case.
The classification performance of the baseline approaches (grey bars) and of the best model per classifier type is presented in Fig.~\ref{fig:AD_full}. 
Firstly, we observe that though the WSLabels baseline outperforms the other baselines in both MA AD datasets, it has poor performance in MA1 AD, which contains very few articles with WS labels for the labels of interest, but performs better in MA2 AD where more WS labels are available for the fine-grained classes.  
% A first observation on these results is that, the WSLabels baseline performs well in MA2, which contains many articles of the narrower classes, and has lower performance in MA1 where WS annotations for the four labels of interest are scarcer. 
% In particular, it seems that WSLabels has better precision than recall.
This is an indication that the heuristic of the \textit{c\textsubscript{i}} occurrence is a good estimation of fine-grained subject labels, when available. Yet it is not satisfactory for more general cases, where the WS labels for the narrower fine-grained are scarce.     
% This suggests that the \textit{c\textsubscript{i}} occurrence is a good heuristic for fine-grained semantic indexing, when available, but it is probably not sufficient for the general case where the narrower concepts are rare. 

% \begin{figure}[htbp]%figure8
% % \centerline{\includegraphics[width=.48\textwidth]{figures/MA1&MA2Best.png}
% \centerline{\includegraphics[width=1\textwidth]{figures/MA1&MA2Best_journal.png}
% }
% \caption{The performance of the baselines (gray) and the best model per classifier type using lexical features only (green) or both lexical and semantic features (blue) on A) The randomly selected MA1 dataset and B) The balanced MA2 dataset. 
% The model is named by the type of the classifier and, inside the parenthesis, the number of selected features and the feature selection method separated by a comma are mentioned, where Heur. stands for heuristic selection. The F1 measure is macro-averaged over the four labels considered (PD, FAD, EOAD and LOAD).
% % The Chi2 and F suffixes in classifier names indicate the feature selection method employed for each model.
% }\label{fig:MA1&MA2Best}
% \end{figure}

When training with lexical features only, the best performing models of all classifier types (green bars in Fig.~\ref{fig:AD_full}) perform better than the baselines in MA1 AD. Some of them using just five lexical features.
% All the best performing models trained with lexical features only (green bars in Fig.~\ref{fig:AD_full}), manage to outperform the baselines in the MA1 AD dataset, regardless of the learning algorithm. Some of them with as few as five lexical features.
This observation supports the hypothesis that models developed by training on WS labels can improve upon the weak-labelling heuristic alone. 
% This fact suggests that models trained on the WS training dataset can improve upon the heuristic employed for weak labelling. 
These models can incorporate missing synonyms or abbreviations not included in the resources of the concept extraction tool. 
This is particularly important for the random MA1 AD dataset where articles with \textit{c\textsubscript{i}} occurrence are rare and WSLabels has poor performance.

On the other hand, only the best models based on logistic regression (LRC) and linear SVC (LSVC) manage to perform close to the best baseline (WSLabels) in the MA2 AD dataset.
% In the MA2 AD dataset, only the best models based on linear SVC (LSVC) and logistic regression (LRC) manage to have a performance close to the strong baseline WSLabels. 
As this dataset was selected using \textit{c\textsubscript{i}} occurrence, most articles have some WS labels and the WSLabels baseline outperforms all the models based on lexical features only. 
This suggests that \textit{c\textsubscript{i}} occurrence can be useful and provide an advantage to the baselines in some cases. 

Considering both types of features, lexical and semantic (blue bars in Fig.~\ref{fig:AD_full}), we observe that random forests (RFC) and decision trees (DTC) perform almost identical to the WSLabels baseline in both datasets. 
% This indicates that they learn to trust the semantic features (the \textit{c\textsubscript{i}} occurrence), which are perfectly correlated to the WS labels they have to learn. 
This suggests that these models learn to trust the \textit{c\textsubscript{i}} occurrence semantic features, which are identical to the WS labels.
% On the other hand, the best performing of these models, which are based on logistic regression (LRC), manage to outperform the baselines in both datasets using both semantic and lexical features. 
However, the logistic regression (LRC) models using both semantic and lexical features, manage to outperform the WSLabels baseline in both MA AD datasets.
% most notably on the MA1 dataset.

\begin{figure}[!t]
\centerline{\includegraphics[width=0.8\textwidth]{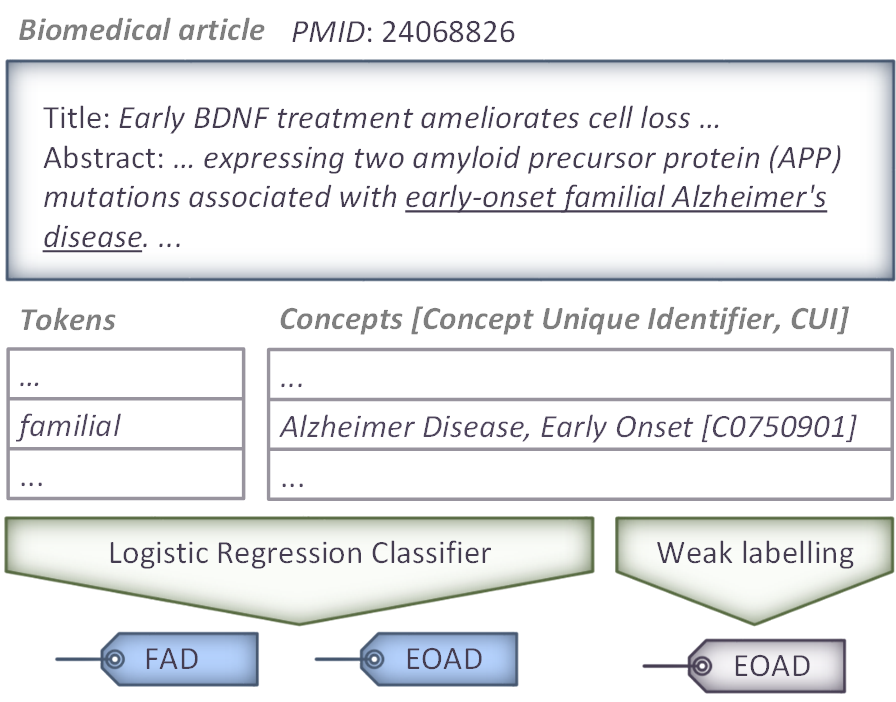}
}
\caption{The fine-grained labels assigned to an article by weak labelling (gray) and by the best Logistic Regression Classifier (LRC) model (blue). 
}\label{fig:PrecitionExample}
\end{figure}

Looking closer at the predictions of LRC models (available in \ref{sec:appendix}), we observe that they also trust the WSLabels baseline to some extent, notably for positive predictions. 
In particular, out of the 58 disagreements of the best LRC models with the WSLabels baseline in the two MA AD datasets, only 6 are due to the LRC model not trusting some WS label.
However, they usually also predict some additional labels, based on features not available to the baselines, leading to improved recall. 
For example, in the article of Fig.~\ref{fig:PrecitionExample}, only the EOAD occurrence is recognised by MetaMap and therefore the WSLabels only predict the EOAD label, missing the FAD label which should also be assigned. However, the LRC model, uses the lexical feature \textit{familial} and manages to predict the FAD label too.

\begin{figure}[!t]
\centerline{\includegraphics[width=1\textwidth]{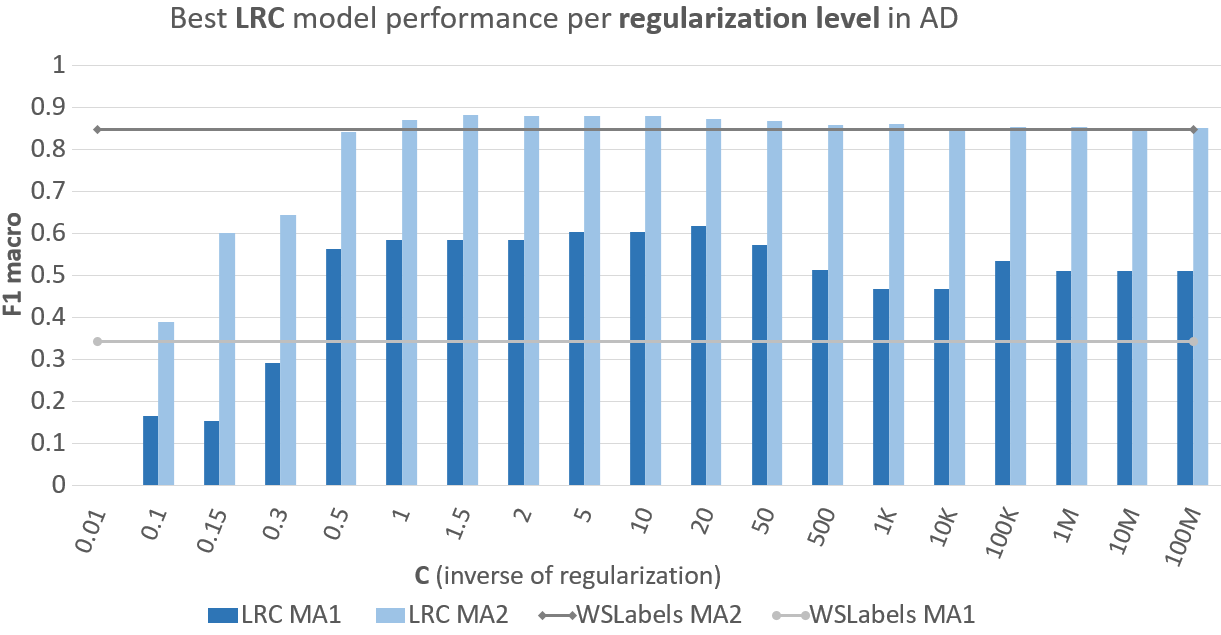}}
\caption{The performance of the best LRC model under different L2 regularization levels in the MA AD datasets. The best performing model is presented for each value of the regularization parameter C. The F1 measure is macro-averaged over the four labels considered (PD, FAD, EOAD and LOAD).
% The Chi2 and F suffixes in classifier names indicate the feature selection method employed for each model.
}\label{fig:RegLevels}
\end{figure}

But what prevents the models based on LRC from being as biased towards the \textit{c\textsubscript{i}} occurrence as the models based on RFC and DTC? 
% This is due to the L2 type regularization performed in the reported experiments, which prevents the model from having extremely high coefficients on only a few features, ignoring all the rest. 
This is because the LRC models are trained with L2 type regularization in these experiments, that prevents them from assigning too high coefficients on just a few features, disregarding all the rest. 
% Experiments with different levels of L2 regularization support this hypothesis. 
Experimentation with LRC models trained under a range of L2 regularization levels supports this hypothesis. 
In Fig.~\ref{fig:RegLevels}, we observe that as the level of L2 regularization becomes lower, that is for higher values of parameter C, the performance of the best LRC model gets closer to the WSLabels baseline, especially in the  MA2 AD dataset.

\begin{figure}[!t]
\centerline{\includegraphics[width=1\textwidth]{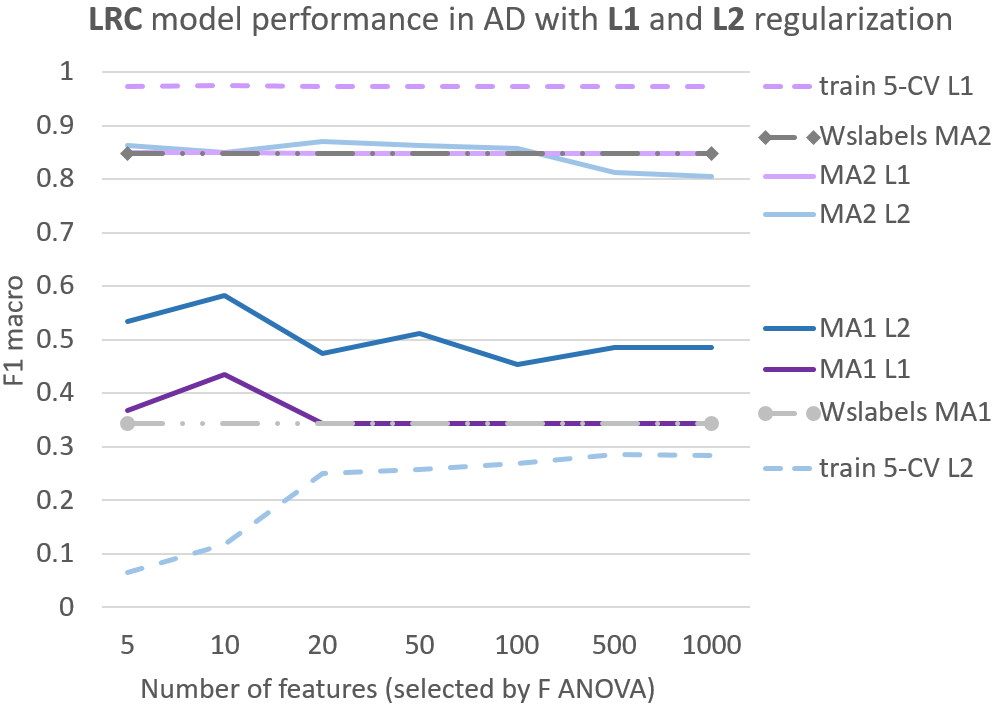}}
\caption{Performance of LRC models trained on the WS AD dataset considering different numbers of features, both lexical and semantic, selected by F ANOVA. The blue and purple lines correspond to models trained with L2 and L1 regularization respectively. The dashed lines show the performance measured with five-fold cross-validation (5-CV) on the WS dataset, while the solid ones the performance on the MA datasets. The dashed grey lines show the performance of the WSLabels baseline on the MA AD datasets. The F1 measure is macro-averaged over the four labels considered (PD, FAD, EOAD and LOAD).
% The Chi2 and F suffixes in classifier names indicate the feature selection method employed for each model.
}\label{fig:RegTypes}
\end{figure}

In addition, experiments with five-fold cross-validation on the training weak supervision dataset, presented in Fig.~\ref{fig:RegTypes}, reveal that under L2 regularization, the LRC models achieve very low cross-validation performance on the training dataset, failing to perform a simple reproduction of the weak labels (dashed blue line). However, when evaluated on the MA AD datasets (solid blue lines), the models perform much better and under some configurations outperform the baseline (grey lines) in both datasets. 

On the other hand, the corresponding LRC models trained with L1 regularization (purple lines), achieve very high cross-validation performance (dashed purple line), but their performance in the MA AD datasets (solid purple lines) is almost identical to the performance of the baseline for most configurations. Moreover, inspection of the coefficients assigned to each feature confirms that the L1 LRC models base their predictions almost exclusively on the corresponding \textit{c\textsubscript{i}} occurrence features. 

Another way to avoid models that just reproduce the weak labels is to explicitly exclude the \textit{c\textsubscript{i}} occurrence features, from the datasets. Experiments for AD under this configuration (orange in Fig.~\ref{fig:AD_full})
% Fig.~\ref{fig:NoLabelFeatures} 
suggest that, even though some models present performance improvements in the random MA1 AD dataset, all of them have lower performance than the WSLabels baseline in the MA2 AD dataset. Therefore, excluding the \textit{c\textsubscript{i}} occurrence features is not the best way to remove the bias that they introduce. 

% \begin{figure}[htbp]%NoLabelFeatures
% % \centerline{\includegraphics[width=.48\textwidth]{figures/MA1&MA2Best.png}
% \centerline{\includegraphics[width=1\textwidth]{figures/NoLabelFeatures_NoUnderSampling.png}}
% \caption{The performance of the best model per classifier type using either all lexical and semantic features (blue) or all lexical and semantic features excluding the semantic ones that correspond to c\textsubscript{i} occurrence (orange) on A) The randomly selected MA1 dataset and B) The balanced MA2 dataset. 
% The model is named by the type of the classifier and, inside the parenthesis, the number of selected features and the feature selection method separated by a comma are mentioned, where Heur. stands for heuristic selection. The F1 measure is macro-averaged over the four labels considered (PD, FAD, EOAD and LOAD).
% }\label{fig:NoLabelFeatures}
% \end{figure} 

As the WS AD dataset is highly skewed, with more than 97\% of the articles annotated with the \textit{c\textsubscript{pref}} label (AD), we investigated under-sampling the articles annotated only with the majority label.
In particular, we experimented with different levels of random \textit{c\textsubscript{pref}} under-sampling with the total size of the majority class ranging from 5,000 articles to a minimum of 752 articles. This minimum corresponds to the case where all articles annotated only with the majority class are removed from the dataset. 
Articles that are labeled with at least one class, apart from \textit{c\textsubscript{pref}}, were retained.
% with any no-majority label, were not removed from the dataset, even if annotated with the majority label too, to avoid loss of information for the labels of interest.
For comparison, the total size of the majority class in the complete WS dataset without under-sampling is 50,111 articles.

% Alternative LRC models where trained on these datasets, similarly to previous experiments.
% and were evaluated on the MA1 and MA2 datasets. 
The results of the under-sampling experiment suggest that some levels of under-sampling can lead to improved predictive performance in the MA2 AD dataset, while performance does not drop in the MA1 AD dataset.
% In particular, in the MA1 dataset most under-sampling experiments lead to best LRC models of performance similar the original dataset. On the other hand, certain levels of under-sampling lead to better predictive performance in MA2 dataset. 
% In particular, a dataset (WS\textsubscript{und}) with 3,000 articles with AD labels lead to improvement of 0.002 and 0.034 in the performance of the best LRC model on the MA1 and MA2 AD datasets respectively. 
The performance of the best models of all four types trained on the WS\textsubscript{und} AD are presented in Fig.~\ref{fig:AD_full} (red bars). 
% Based on these experiments, we adopted the WS\textsubscript{und} dataset for further experiments, and the performance of the best models of all four types trained on the WS\textsubscript{und} are presented in Fig.~\ref{fig:WSundResults} (red bars)
These results suggest that apart from the LRC models, other types of models also achieve small performance improvements when trained on this more balanced, under-sampled WS\textsubscript{und} dataset. The distribution of labels in WS\textsubscript{und} is presented in Table~\ref{tabADinitial}.

Since the LRC models predict better than the \textit{c\textsubscript{i}} occurrence heuristic on the MA AD datasets, it is interesting to investigate whether the predictions of these models could also be exploited to train new models that further improve the predictions.
In this direction, we assigned to the articles of the  WS\textsubscript{und} AD dataset the labels predicted by the best  LRC models on the MA AD datasets. Then, we trained new LRC models on this re-labeled version of the WS\textsubscript{und} AD dataset and evaluated the new predictions in the MA AD datasets. The results revealed no improvement and in some cases a drop in classification performance.
% [todo]
% \todo{Explanation for this?}

% If the best performing LRC models on the MA datasets are indeed capable to predict better annotations than the concept-occurrence heuristic, in other datasets too, it is interesting to investigate whether training new models based on the predicted labels for the articles of the WS\textsubscript{und} dataset could further improve the predictions. 
% In this direction, Fig.~\ref{fig:Iterations} presents some experiments where the labels predicted by the best LRC models trained on WS\textsubscript{und} dataset (red bars) were used for training new LRC models, under the same configuration, iteratively (purple bars). 
% The performance of the models in both MA datasets drops rapidly in each iteration which suggests that the selected models predict misleading labels in the WS\textsubscript{und} dataset. 
% In a similar experiment, presented in Fig.~\ref{fig:IterationsWS} we follow the same procedure with reduced L2 regularization (C=100 instead of 1), to train models with less bias that would fit better to the training data.
% In these experiments, the performance of the models trained on the predicted labels doesn't drop that much. In particular, in the MA2 dataset some marginal improvement is observed for some iterations. However, no model has performance as good as the initial model trained with C=1 in the MA datasets, even though they are really close.
% \todo[inline]{"error analysis" e.g. snippets from articles missed by WS but found by LGC?}

\subsubsection{Results on Duchenne Muscular Dystrophy}
\begin{figure}[!tp]%figure8
% \centerline{\includegraphics[width=.48\textwidth]{figures/MA1&MA2Best.png}
\centerline{\includegraphics[width=1\textwidth]{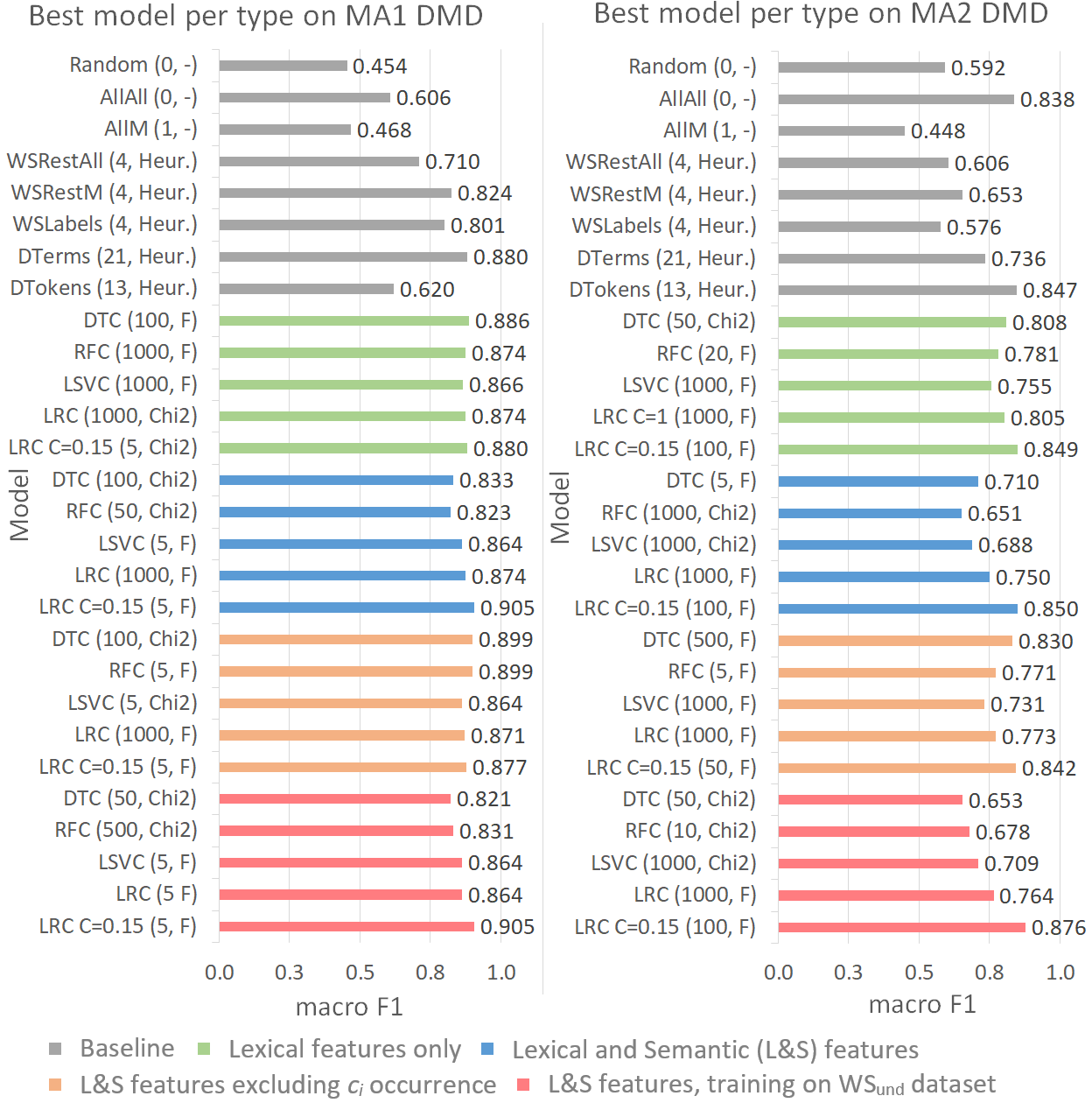}
}
\caption{The performance of the baselines (gray) and the best model per classifier type assessed on the randomly selected MA1 DMD dataset (left) and the balanced MA2 DMD dataset (right). For each classifier type, models are trained on the WS dataset using lexical features only (green), both lexical and semantic (L\&S) features (blue) and lexical and semantic features, excluding the semantic ones that correspond to \textit{c\textsubscript{i}} occurrence (orange). In addition models using both lexical and semantic features have also been trained in the WS\textsubscript{und} dataset (red). 
Each model is named by the type of the classifier and inside a parenthesis, the number of selected features and the feature selection method separated by a comma. Heur. stands for heuristic selection. 
For the LRC models, apart from the default L2 regularization level, that is C=1, models with increased L2 regularization are also assessed, which is denoted with LRC C=0.15.
The F1 measure is macro-averaged over the two labels considered (DMD and BMD).
}\label{fig:DMD_full}
\end{figure}

The results on the DMD use case, presented in Fig.~\ref{fig:DMD_full} differ in many ways to those of the AD use case. Firstly, both the MA DMD datasets seem less challenging than the AD ones, as random and trivial baselines achieve much higher macro-F1 performance. 
This is mostly due to the high prevalence of the DMD concept, which is relevant to more than 80\% of the articles even in the ``balanced'' MA2 DMD dataset. Contrary to AD, in this use case the \textit{c\textsubscript{pref}} (DMD) is other than the  \textit{c\textsubscript{top}} (DBMD) and is therefore considered for fine-grained labelling, leading to higher imbalance of the labels to predict.   

Similar to AD, WSLabels is a strong heuristic in the randomly selected MA1 DMD, but has a macro-F1 performance close to random in the MA2 DMD, where the trivial baseline AllAll achieves a high performance above 0.8 macro-F1. 
This is due to low recall of the WSLabels for the abundant DMD label which is relevant to 81 out of 100 articles. Although WSLabels achieves high precision for both BMD and DMD, it is penalised for assigning the DMD label only to 26 articles out of the 81, achieving a recall as low as 0.31 for this label, when Random achieves 0.54. 
% It seems that this low recall of WSLabels is related to the recognition of the broader DBMD concept instead of the DMD, which highlights the importance of   
This suggests that concept-occurrence is precise at detecting concept-specific articles, but it can lose in recall. 

On the other hand, the dictionary-based baselines perform really well in this use case with DTerms and DTokens achieving the best baseline performance in MA1 and MA2 respectively.
The high coverage of MeSH in synonymous terms for the concepts of this use-case, providing 15 and 6 terms for DMD and BMD respectively compared to only 7 terms for all four labels of the AD use case,
% \footnote{There are 15 and 6 terms for DMD and BMD respectively, compared to just 3, 2, 1 and 1 terms for EOAD, LOAD, FAD and PD.}
 may contribute in this high performance of dictionary-based approaches.  

The best models trained on WS DMD considering lexical features only (green in Fig.~\ref{fig:DMD_full}) perform better than the WSLabels baseline in both MA DMD datasets but can't outperform the strongest dictionary-based baselines. 
Training LRC models with more L2 regularisation (C=0.15 instead of the default C=1) leads to improvements in the best LRC models, which eventually manage to perform no worse than the best baselines in both datasets. 

With the addition of semantic features (blue in Fig.~\ref{fig:DMD_full}) most models achieve lower performance, closer to the WS baselines in both MA DMD datasets. However, the properly regularised LRC models manage to take advantage of the semantic features and further improve their classification performance, though marginally, as has been observed in the AD MA2 dataset too.

Removal of the \textit{c\textsubscript{i}} occurrence features (orange in Fig.~\ref{fig:DMD_full}) leads to performance improvements in some models for both MA DMD datasets, without exceeding the performance of the best model using lexical and semantic features. This suggests, that regardless of the use case, semantic features, with \textit{c\textsubscript{i}} occurrence included, can be useful for fine-grained semantic indexing, under proper L2 regularisation.

Under-sampling experiments in the DMD use case (red in Fig.~\ref{fig:DMD_full}), also confirm that some levels of balancing the WS dataset can benefit the classification performance. In particular, training on a dataset (WS\textsubscript{und} DMD) with 1,000 articles with DMD labels, leads the best LRC models with increased L2 regularisation to higher performance in MA2 DMD, without affecting the performance in MA1 DMD. 

Finally, experiments with re-labelling and re-training did not lead to higher performance, similar to the AD use case.
% the WS\textsubscript{und} DMD based on the best predictions on the MA DMD datasets and training new models, had similar results as in the  AD use case, without any improvement achieved.

\section{Conclusion and discussion}
\label{sec:Conclusion}
The main contributions of this paper comprise the formulation of concept-level semantic indexing as a multi-label classification task, the proposal of a new method for automated development of weakly supervised predictive models for this problem and the assessment of the new method in two real use cases, about two different diseases, to investigate and demonstrate its feasibility. 
% The contribution of this paper is focused on formulating the fine-grained semantic indexing problem as a multi-label classification task, suggesting a method to automatically produce weakly supervised classifiers for the task and demonstrating the feasibility of applying this method in two real use cases. 

Particularly, our results suggest that the weak labeling based on the concept occurrence is a strong heuristic for concept-level fine-grained semantic indexing.
% In particular, we show that heuristic labeling of articles with concept occurrence is a good estimation for fine-grained semantic labels, though far from perfect. 
Additionally, it seems that models trained on the weakly labelled data can outperform the heuristic baselines under some configurations, providing better fine-grained subject annotations.
% We also present some models that manage to outperform the strong baseline in some cases, suggesting that training with weak labels based on concept occurrence, can produce predictive models that can indeed generalize and produce annotations better than concept occurrence itself. 

In addition, we experimented with the use of different semantic features in the predictive models, highlighting issues related to the perfect correlation of certain features with the heuristic labels used as weak supervision.
These features seem useful, but they can also bias and misguide some classifiers. L2 regularization seemed to remove this bias and allow the logistic regression classifiers to achieve very good results in the specific use cases.
Further experiments with a range of under-sampling levels suggest that balancing the training dataset can have beneficial effects in model performance. Experimentation with iterative training of models on predicted labels 
% revealed that this setting performs better under lower L2 regularization levels, but 
didn't result in overall improvement. 
% These initial results indicate that there is still improvement potential under certain configurations.

In conclusion, our results on two real use-cases, though not sufficient for safe generalisation to any use-case, suggest that using concept-occurrence as weak supervision for fine-grained semantic indexing is feasible, and training Logistic regression models with L2 regularization leads to the best models. Both lexical and semantic features are useful, including the ones used for weak labelling, and usually no more than one hundred features are needed. Finally, under-sampling the \textit{c\textsubscript{pref}} label, to keep just a few thousands of instances, can also benefit the models.
 
Our future plans include the application of the proposed method in a variety of diseases to support more general conclusions and eventually lead into a system sufficiently comprehensive for direct usage into new use-cases.
In addition, we aim at improving the predictive performance of the classification models further and plan to investigate the extension of the set of concept-level labels, integrating more concepts from UMLS vocabularies or even emerging concepts not yet in the vocabularies.
% fine-tunig the parameters of each model and employing alternative overfitting avoidance techniques on each algorithm to maximize the ability of the models to generalize, providing fine-grained semantic labels of improved quality for the the biomedical literature.     

Our motivation is to support a new search mechanism that will exploit the automated fine-grained annotations providing to the users targeted access to biomedical literature for a specific topic of their expertise, like a disease sub-type. For instance looking for articles relevant to early-onser Alzheimer's disease (EOAD), we aim at search results with better balance of precision and recall, than searching with the MeSH descriptor for AD or with the terms of the EOAD concept.

\section{Acknowledgment}

This work was partially supported by the EU H2020 programme, under grant agreement No 727658 (project iASiS). We are grateful to Ingrid Verhaart, Nikil Patel, Natasha Clarke and Peter Garrard for kindly contributing in the manual annotation of the datasets for testing the proposed method.

% \section{Front matter}

% The author names and affiliations could be formatted in two ways:
% \begin{enumerate}[(1)]
% \item Group the authors per affiliation.
% \item Use footnotes to indicate the affiliations.
% \end{enumerate}
% See the front matter of this document for examples. You are recommended to conform your choice to the journal you are submitting to.

% \section{Bibliography styles}

% There are various bibliography styles available. You can select the style of your choice in the preamble of this document. These styles are Elsevier styles based on standard styles like Harvard and Vancouver. Please use Bib\TeX\ to generate your bibliography and include DOIs whenever available.

% Here are two sample references: \cite{Feynman1963118,Dirac1953888}.

\appendix

\section{WS and LRC label disagreements in the AD use case}
\label{sec:appendix}
The attached Excel file, presents the disagreements between  WS and predicted labels in MA AD datasets. In particular, the predictions of the best LRC model trained on WS AD dataset considering lexical and semantic features (blue bars in Fig.~\ref{fig:AD_full}) are compared to the best baseline (WSLabels). In addition, the values of the features are also presented for each article, as well as the coefficients of the corresponding LRC model for each feature. Articles where the predicted and the WS labels are in total agreement are omitted from these tables.

\section*{References}

\bibliography{mybibfile}

\end{document}